%%%%%%%%%%%%%%%%%%%%%%%%%%%%%%%%%%%%%%%%%%%%%%%%%%%%%%%%%%%%%%%%%%%%
%  TeX Definitions                                                 %
%%%%%%%%%%%%%%%%%%%%%%%%%%%%%%%%%%%%%%%%%%%%%%%%%%%%%%%%%%%%%%%%%%%%
\input harvmac
\noblackbox
%%% Figures
\newcount\figno
\figno=0
\def\fig#1#2#3{
\par\begingroup\parindent=0pt\leftskip=1cm\rightskip=1cm\parindent=0pt
\baselineskip=11pt

\global\advance\figno by 1
\midinsert
\epsfxsize=#3
\centerline{\epsfbox{#2}}
\vskip 12pt
\centerline{{\bf Figure \the\figno:} #1}\par
\endinsert\endgroup\par}
\def\figlabel#1{\xdef#1{\the\figno}}

\def\np#1#2#3{Nucl. Phys. {\bf B#1} (#2) #3}
\def\pl#1#2#3{Phys. Lett. {\bf B#1} (#2) #3}
\def\prl#1#2#3{Phys. Rev. Lett.{\bf #1} (#2) #3}
\def\physrev#1#2#3{Phys. Rev. {\bf D#1} (#2) #3}

%%% Paragraphs

%%% special math symbols
\font\cmss=cmss10
\font\cmsss=cmss10 at 7pt
\def\rlx{\relax\leavevmode}
\def\inbar{\vrule height1.5ex width.4pt depth0pt}
\def\IC{\relax\,\hbox{$\inbar\kern-.3em{\rm C}$}}
\def\IN{\relax{\rm I\kern-.18em N}}
\def\IP{\relax{\rm I\kern-.18em P}}
\def\ZZ{\rlx\leavevmode\ifmmode\mathchoice{\hbox{\cmss Z\kern-.4em Z}}
  {\hbox{\cmss Z\kern-.4em Z}}{\lower.9pt\hbox{\cmsss Z\kern-.36em Z}}
  {\lower1.2pt\hbox{\cmsss Z\kern-.36em Z}}\else{\cmss Z\kern-.4em
  Z}\fi}
%%% misc.
\def\IZ{\relax\ifmmode\mathchoice
{\hbox{\cmss Z\kern-.4em Z}}{\hbox{\cmss Z\kern-.4em Z}}
{\lower.9pt\hbox{\cmsss Z\kern-.4em Z}}
{\lower1.2pt\hbox{\cmsss Z\kern-.4em Z}}\else{\cmss Z\kern-.4em
Z}\fi}

\def\narrowplus{\kern -.04truein + \kern -.03truein}
\def\narrowminus{- \kern -.04truein}
\def\narrowminussub{\kern -.02truein - \kern -.01truein}

\def\kh{K\"{a}hler }
\def\half{{1\over 2}}

\def\b{{\beta}}
\def\a{{\alpha}}
\def\g{{\gamma}}

\def\vol{{\rm vol}}

\def\r{{\rightarrow}}

\def\frac#1#2{{#1\over #2}}

\def\CM{{\cal M}}

\def\BR{\IR}
\def\BZ{\IZ}

\def\IZ{\relax\ifmmode\mathchoice
{\hbox{\cmss Z\kern-.4em Z}}{\hbox{\cmss Z\kern-.4em Z}}
{\lower.9pt\hbox{\cmsss Z\kern-.4em Z}}
{\lower1.2pt\hbox{\cmsss Z\kern-.4em Z}}\else{\cmss Z\kern-.4em
Z}\fi}
\def\IB{\relax{\rm I\kern-.18em B}}
\def\IC{{\relax\hbox{$\inbar\kern-.3em{\rm C}$}}}
\def\ID{\relax{\rm I\kern-.18em D}}
\def\IE{\relax{\rm I\kern-.18em E}}
\def\IF{\relax{\rm I\kern-.18em F}}
\def\IG{\relax\hbox{$\inbar\kern-.3em{\rm G}$}}
\def\IGa{\relax\hbox{${\rm I}\kern-.18em\Gamma$}}
\def\IH{\relax{\rm I\kern-.18em H}}
\def\II{\relax{\rm I\kern-.18em I}}
\def\IK{\relax{\rm I\kern-.18em K}}
\def\IP{\relax{\rm I\kern-.18em P}}
%\def\IX{\relax{\rm X\kern-.01em X}}
%this doesn't work

\font\cmss=cmss10 \font\cmsss=cmss10 at 7pt
\def\IR{\relax{\rm I\kern-.18em R}}

\def\S{{\Sigma}}

%

%
%       \eqn\label{a+b=c}	gives displayed equation, numbered
%				consecutively within sections.
%     \eqnn and \eqna define labels in advance (of eqalign?)
%
\def\eqnn#1{\xdef #1{(\secsym\the\meqno)}\writedef{#1\leftbracket#1}%
\global\advance\meqno by1\wrlabeL#1}
\def\eqna#1{\xdef #1##1{\hbox{$(\secsym\the\meqno##1)$}}
\writedef{#1\numbersign1\leftbracket#1{\numbersign1}}%
\global\advance\meqno by1\wrlabeL{#1$\{\}$}}
\def\eqn#1#2{\xdef #1{(\secsym\the\meqno)}\writedef{#1\leftbracket#1}%
\global\advance\meqno by1$$#2\eqno#1\eqlabeL#1$$}

%%

%%%%%%%%%%%%%%%%%%%%%%%%%%%%%%%%%%%%%%%%%%%%%%%%%%%%%%%%%%%%%%%%%%%%%%%%%%%%
%                    My definitions                                        %
%%%%%%%%%%%%%%%%%%%%%%%%%%%%%%%%%%%%%%%%%%%%%%%%%%%%%%%%%%%%%%%%%%%%%%%%%%%%
\input epsf

\def\SUSY#1{{{\cal N}= {#1}}}                   % N=? SUSY
\def\lbr{{\lbrack}}                             % [
\def\rbr{{\rbrack}}                             % ]

\def\wdg{{\wedge}}                              % wedge product

                              % inverse
                           % O(x)

\def\MR#1{{{\BR}^{#1}}}               % Real numbers
               % Complex numbers

\def\MR#1{{{\BR}^{#1}}}               % Real numbers
               % Complex numbers
\def\MS#1{{{\bf S}^{#1}}}               % Circle, sphere,...
               % disk, ball,...
\def\MT#1{{{\bf T}^{#1}}}               % Torus
               % CP
               % Ruled surface F_n

             % Patch
                    % line-bundle
            % line-bundle of a divisor
\def\px#1{{\partial_{#1}}}              % derivative
\def\pypx#1#2{{{\partial {#1}}\over {\partial {#2}}}}      % derivative

                 % Left large bracket
                % Right large bracket
              % SL(*,Z)

                             % identity matrix

      % commutator
               % anti-commutator

           % expectation value

    % expectation value of trace

      % trace
\def\trp#1{{{\rm tr}\{ {#1} \} }}            % trace
            % Trace
            % trace in a rep
            % Trace in a rep

\def\rep#1{{{\bf {#1}}}}                      % representation
                  % Imaginary
                  % Real

%\def\widebar#1{{\bar{#1}}}                    % Wide bar
\def\widebar#1{{\overline{#1}}}                    % Wide bar

\def\MHT#1{{\widehat{{\bf T}}^{#1}}}               % Torus

 % dual torus

%%% \def\subset{{{<}}}   % group subset
\mathchardef\subset="321A

\def\pV{{p_\Vert}}
\def\RV{{R_\Vert}}
      % The matrix model coupling constant
\def\MHWT#1{{{\widehat{\widetilde{\bf T}}}{}^{#1}}}
\def\wM{{\widetilde{M}}}
\def\MWT#1{{ {\bf \widetilde{T}}^{#1} }}  % auxiliary T^2

\def\u{{\mu}}

\def\lam{{\lambda}}

\def\X{{ \cal C}}

%%%%%%%%%%%%%%%%%%%%%%%%%%%%%%%%%%%%%%%%%%%%%%%%%%%%%%%%%%%%%%%%%%%%
%  B I B L I O G R A P H Y                                         %
%%%%%%%%%%%%%%%%%%%%%%%%%%%%%%%%%%%%%%%%%%%%%%%%%%%%%%%%%%%%%%%%%%%%

\lref\rpol{J. Polchinski, ``TASI Lectures on D-Branes,''
    hep-th/9611050\semi J. Polchinski, S. Chaudhuri and C. Johnson,
    ``Notes on D-Branes,'' hep-th/9602052. }

\lref\rBFSS{T. Banks, W. Fischler, S. H. Shenker, and L. Susskind, ``M
    Theory As A Matrix Model: A Conjecture,''
    hep-th/9610043, Phys. Rev. {\bf D55} (1997) 5112.}

\lref\rwtensor{E. Witten, ``Some Comments on String Dynamics,''
    hep-th/9507121.}

\lref\rstensor{A. Strominger, ``Open P-Branes,'' hep-th/9512059,
    \pl{383}{1996}{44}.}

\lref\rsdecoupled{N. Seiberg, ``New Theories in Six Dimensions and
    Matrix Description of M-theory on $T^5$ and $T^5/\IZ_2$,''
    hep-th/9705221.}

\lref\rashoke{ A. Sen, ``A Note on Enhanced Gauge Symmetries in
    M and String Theory,'' hep-th/9707123; ``Dynamics of Multiple
    Kaluza-Klein Monopoles in M and String Theory,'' hep-th/9707042.}

\lref\kutetal{D. Berenstein, R. Corrado and J. Distler, ``On the
    Moduli Spaces of M(atrix)-Theory Compactifications,''
    hep-th/9704087\semi  S. Elitzur, A. Giveon, D. Kutasov and
    E. Rabinovici, ``Algebraic Aspects of Matrix Theory on $T^d$,''
    hep-th/9707217.}

\lref\rtwoform{J. P. Gauntlett and D. Lowe, ``Dyons and S-Duality in
    N=4 Supersymmetric Gauge Theory,'' hep-th/9601085,
    \np{472}{1996}{194}\semi K. Lee, E. Weinberg and P. Yi,
    ``Electromagnetic Duality and $SU(3)$ Monopoles,'' hep-th/9601097,
    \pl{376}{1996}{97}.}

\lref\rmoore{A. Losev, G. Moore, and S. Shatashvili, ``M \& m's ,''
    hep-th/9707250.}

\lref\rbrunner{I. Brunner and A. Karch, ``Matrix Description of
    M-theory on $T^6$,'' hep-th/9707259.}

\lref\rSeiWHY{N. Seiberg, ``Why is the Matrix Model Correct?,''
hep-th/9710009.}

\lref\rSenTn{A. Sen,
  ``D0 Branes on $T^n$ and Matrix Theory,'' hep-th/9709220.}

\lref\rDVV{R. Dijkgraaf, E. Verlinde and H. Verlinde, ``BPS Spectrum
    of the Five-Brane and Black Hole Entropy,'' hep-th/9603126,
    \np{486}{1997}{77}; ``BPS Quantization of the Five-Brane,''
    hep-th/9604055, \np{486}{1997}{89}.}

\lref\rextraDVV{R. Dijkgraaf, E. Verlinde and H. Verlinde, ``5D Black Holes and
Matrix Strings,'' hep-th/9704018.}

\lref\rsixbrane{P. Townsend, ``The Eleven Dimensional Supermembrane
    Revisited,'' hep-th/9501068, \pl{350}{1995}{184}.}

\lref\rmultitn{R. Sorkin, ``Kaluza-Klein Monopole,''
    \prl{51}{1983}{87}\semi D. Gross and M. Perry, ``Magnetic Monopoles
    in Kaluza-Klein Theories,'' \np{226}{1983}{29}.}

\lref\rmIIB{P. Aspinwall, ``Some Relationships Between Dualities in
    String Theory,'' hep-th/9508154, Nucl. Phys. Proc. Suppl. {\bf 46}
    (1996) 30\semi J. Schwarz, ``The Power of M Theory,''
    hep-th/9510086, \pl{367}{1996}{97}. }

\lref\rgeneralsixbrane{ C. Hull,  ``Gravitational Duality, Branes and
    Charges,'' hep-th/9705162\semi E. Bergshoeff, B. Janssen, and
    T. Ortin, ``Kaluza-Klein Monopoles and Gauged Sigma Models,''
    hep-th/9706117\semi Y. Imamura, ``Born-Infeld Action and
Chern-Simons
    Term {}from Kaluza-Klein Monopole in M-theory,'' hep-th/9706144.}

\lref\rbd{M. Berkooz and M. Douglas, ``Five-branes in M(atrix)
    Theory,'' hep-th/9610236, \pl{395}{1997}{196}.}

\lref\rbraneswith{M. Douglas, ``Branes within Branes,''
    hep-th/9512077.}

\lref\rquantumfive{O. Aharony, M. Berkooz, S. Kachru, N. Seiberg, and
    E. Silverstein, ``Matrix Description of Interacting Theories in Six
    Dimensions,'' hep-th/9707079.}

\lref\rstringfive{E. Witten, ``On The Conformal Field Theory of The
    Higgs Branch,'' hep-th/9707093.}
    \lref\rSS{S. Sethi and L. Susskind, ``Rotational Invariance in the
    M(atrix) Formulation of Type IIB Theory,'' hep-th/9702101,
    \pl{400}{1997}{265}.}

\lref\rBS{T. Banks and N. Seiberg, ``Strings from Matrices,''
    hep-th/9702187, \np{497}{1997}{41}.}

\lref\rgilad{A. Hanany and G. Lifschytz, ``M(atrix) Theory on $T^6$
    and a m(atrix) Theory Description of KK Monopoles,''
    hep-th/9708037.}

\lref\rreview{N. Seiberg, ``Notes on Theories with 16
    Supercharges,'' hep-th/9705117.}

\lref\rDVVstring{R. Dijkgraaf, E. Verlinde and H. Verlinde, ``Matrix
    String Theory,'' hep-th/9703030.}
    \lref\rprobes{M. Douglas, ``Gauge Fields and D-branes,''
    hep-th/9604198.}

\lref\rdoumoo{M. Douglas and G. Moore, ``D-Branes,
    Quivers, and ALE Instantons,'' $\,$ hep-th/9603167\semi
    C. Johnson and R. Myers, ``Aspects of Type IIB Theory on ALE
    Spaces,'' hep-th/9610140, \physrev{55}{1997}{6382}.}

\lref\rdouglas{M. Douglas, ``Enhanced Gauge Symmetry in M(atrix)
    Theory,'' hep-th/9612126\semi W. Fischler and A. Rajaraman,
 ``M(atrix) String Theory on K3,'' hep-th/9704123.}

\lref\rsprobes{N. Seiberg, ``Gauge Dynamics And Compactification To
    Three Dimensions,'' hep-th/9607163, \pl{384}{1996}{81}}

\lref\rswthree{N. Seiberg and E. Witten, ``Gauge Dynamics and
    Compactifications to Three Dimensions,'' hep-th/9607163.}

\lref\rthroat{D.-E. Diaconescu and N. Seiberg, ``The Coulomb Branch of
    $(4,4)$ Supersymmetric Field Theories in Two Dimensions,''
    hep-th/9707158. }

\lref\rTduality{T. Banks, M. Dine, H. Dykstra and W. Fischler,
    ``Magnetic Monopole Solutions of String Theory,''
    \pl{212}{1988}{45}\semi C. Hull and P. Townsend, ``Unity of
    Superstring Dualities,'' hep-th/9410167, \np{438}{109}{1995}\semi
    H. Ooguri, C. Vafa, ``Two Dimensional Black Hole and Singularities
    of Calabi-Yau Manifolds,'' Nucl.Phys. {\bf B463} (1996) 55,
    hep-th/9511164\semi D. Kutasov, ``Orbifolds and Solitons,''
    Phys.  Lett {\bf B383} (1996) 48, hep-th/9512145\semi
    H. Ooguri and C. Vafa,
    ``Geometry of N=1 Dualities in Four Dimensions,'' hep-th/9702180.}

\lref\raps{P. Argyres, R. Plesser and N. Seiberg, ``The Moduli Space
    of N=2 SUSY QCD and Duality in N=1 SUSY QCD,'' hep-th/9603042,
    \np{471}{1996}{159}.}

\lref\rgms{O. Ganor, D. Morrison and N. Seiberg, ``Branes, Calabi-Yau
    Spaces, and Toroidal Compactification of the N=1 Six Dimensional
    $E_8$ Theory,'' hep-th/9610251, \np{487}{1997}{93}.}

\lref\rchs{ C.G. Callan, J.A. Harvey, A. Strominger, ``Supersymmetric
    String Solitons,'' hep-th/9112030, \np{359}{1991}{611}\semi
    S.-J. Rey, in ``The  Proc. of the Tuscaloosa Workshop
    1989,'' 291; Phys. Rev. {\bf D43} (1991) 526; S.-J. Rey, In DPF
    Conf. 1991, 876.}

\lref\rmotl{L. Motl, ``Proposals on nonperturbative superstring
    interactions,'' hep-th/9701025.}

\lref\rwati{W. Taylor IV, ``D-Brane Field Theory on Compact Space,''
    hep-th/9611042, \pl{394}{1997}{283}.}

\lref\rkin{K. Intriligator, ``New String Theories in Six Dimensions
    via Branes at Orbifold Singularities,'' hep-th/9708117.}

\lref\rmont{C. Montonen and D. Olive, \pl{72}{1977}{117}.}

\lref\rsdual{A. Sen, ``Dyon-Monopole Bound States, Self-Dual Harmonic
    Forms on the Multi-Monopole Moduli Space, and SL(2,Z) Invariance in
    String Theory,'' hep-th/9402032, \pl{329}{1994}{217}.}

\lref\rstrong{C. Vafa and E. Witten, ``A Strong Coupling Test of
    S-Duality,'' hep-th/9408074, \np{432}{1994}{3}. }

\lref\rlenny{L. Susskind, ``T Duality in M(atrix) Theory and S Duality
    in Field Theory,'' hep-th/9611164. }

\lref\rori{O. Ganor, S. Ramgoolam and W. Taylor IV, ``Branes, Fluxes and
    Duality in M(atrix)-Theory,'' hep-th/9611202, \np{492}{1997}{191}.}

\lref\rfourtorus{M. Rozali, ``Matrix Theory and U-Duality in Seven
    Dimensions,'' hep-th/9702136, Phys. Lett. {\bf B400} (1997) 260\semi
M, Berkooz, M. Rozali and N. Seiberg, ``Matrix Description of M-theory
on $T^4$ and $T^5$,'' hep-th/9704089.}

\lref\rnati{N. Seiberg and S. Sethi, ``Comments on Neveu-Schwarz
    Five-Branes,'' hep-th/9708085.}

\lref\rdlcq{L. Susskind, ``Another Conjecture about M(atrix) Theory,''
    hep-th/9704080.}

\lref\rsIIB{S. Sethi, ``The Matrix Formulation of Type IIB
    Five-Branes,'' hep-th/9710005.}

\lref\reIIB{E. Witten, ``New `Gauge' Theories in Six-Dimensions,''
    hep-th/9710065.}

\lref\rhmon{S. Sethi and M. Stern, ``A Comment on the Spectrum of
    H-Monopoles,'' hep-th/9607145, \pl{398}{1997}{47}.}

\lref\rconjecture{J. de Boer, K. Hori, H. Ooguri and Y. Oz, ``Mirror
    Symmetry in Three-Dimensional Gauge Theories, Quivers and D-branes,''
    hep-th/9611063, \np{493}{1997}{101}.}

\lref\rmirror{K. Intriligator and N. Seiberg, ``Mirror Symmetry in Three
    Dimensional Gauge Theories,'' hep-th/9607207, \pl{386}{1996}{513}. }

\lref\rBR{J. Brodie and S. Ramgoolam, ``On Matrix Models of M5 branes,''
hep-th/9711001.}

\lref\rjulie{J. D. Blum and K. Intriligator, ``Consistency Conditions for
Branes at
Orbifold Singularities,'' hep-th/9705030; ``New Phases of String Theory and 6d
RG
Fixed Points via Branes at Orbifold Singularities,'' hep-th/9705044.}

\lref\rImamura{Y. Imamura,
   ``A Comment on Fundamental Strings in M(atrix) Theory,''
  hep-th/9703077.}

\lref\rEGKRalg{S. Elitzur, A. Giveon, D. Kutasov, E. Rabinovici,
   ``Algebraic Aspects of Matrix Theory on $T^d$,''
  hep-th/9707217.}

\lref\rGivKut{A. Giveon and D. Kutasov, to appear.}

\lref\rHacVer{F. Hacquebord and H. Verlinde,
   ``Duality symmetry of $\SUSY{4}$ Yang-Mills theory on $T^3$,''
  hep-th/9707179.}

\lref\rWitBR{E. Witten,
  ``Solutions Of Four-Dimensional Field Theories Via M Theory,''
  hep-th/9703166.}

\lref\rtHooft{G. 't Hooft,
  ``On the phase transition towards permanent quark confinement,''
  \np{138}{1978}{1-25}.}

\lref\rHelPol{S. Hellerman and J. Polchinski,
  ``Compactification in the Light-Like Direction,''
  hep-th/9711037.}

\lref\rVafIOD{A. Sen, ``U-duality and Intersecting D-branes,''
hep-th/9511026\semi
 C. Vafa, ``Gas of D-Branes and Hagedorn Density of BPS States,''
hep-th/9511088.}

\lref\rGriHar{P. Griffiths and J. Harris,
  {\it Principles of Algebraic Geometry}, Wiley-Interscience, New York,
  1978.}

\lref\rGukov{S. Gukov,
  ``Seiberg-Witten Solution from Matrix Theory,'' hep-th/9709138.}

\lref\rFMW{R. Friedman, J. Morgan, and E. Witten,
   ``Vector Bundles And F Theory,'' hep-th/9701162\semi
M. Bershadsky, A. Johansen, T. Pantev and V. Sadov, ``On Four-Dimensional
Compactifications of F-Theory, '' hep-th/9701165.}

\lref\rEGKRalg{S. Elitzur, A. Giveon, D. Kutasov, E. Rabinovici,
  ``Algebraic Aspects of Matrix Theory on ${\bf T}^d$,''
  hep-th/9707217.}

\lref\rneqone{S. Elitzur, A. Giveon, D. Kutasov,
  {``Brane and $\SUSY{1}$ Duality In String Theory,''}
  hep-th/9702014.}

%  {``Brane Dynamics and $\SUSY{1}$ Supersymmetric Gauge Theory,''}
%  hep-th/9704104\semi
%  K. Hori, H. Ooguri and Y. Oz,
%  {``Strong Coupling Dynamics of Four-Dimensional
%  $\SUSY{1}$ Gauge Theories from M-Theory Fivebrane,''} hep-th/9706082.}

\lref\rmukai{S. Mukai, ``Duality between $D(X)$ and $D(
  \hat{X}$ ) with its Application to Picard Sheaves,'' Nagoya.
  Math. J. {\bf 81} (1981) 153.}

\lref\rKMV{S. Katz, P. Mayr and C. Vafa, ``Mirror Symmetry and Exact Solution
of 4D
  N=2 Gauge Theories - I,'' hep-th/9706110.}

\lref\rBarak{B. Kol, ``On 6d ``Gauge'' Theories with Irrational Theta Angle, ''
hep-th/
9711017.}

\lref\rDOS{M. Douglas, H. Ooguri and S. Shenker, ``Issues in M(atrix)
 Theory Compactification,'' hep-th/9702203.}

\lref\rBerRoz{S. Govindarajan, ``A note on M(atrix) theory in seven
dimensions with eight supercharges, '' hep-th/9705113\semi
M. Berkooz and M. Rozali, ``String dualities from M(atrix) Theory,''
hep-th/9705175.}

\lref\rSeiFDS{N. Seiberg,
  { ``Five Dimensional SUSY Field Theories, Non-trivial Fixed Points
  And String Dynamics,''} hep-th/9608111.}

\lref\rmayr{A. Klemm, W. Lerche, P. Mayr and C. Vafa, ``Self-Dual
Strings and N=2 Supersymmetric Field Theory,'' hep-th/9604034.}

\lref\ryi{K. Lee and P. Yi, ``Monopoles and Instantons on Partially
Compactified D-Branes,'' hep-th/9702107.}

%%%%%%%%%%%%%%%%%%%%%%%%%%%%%%%%%%%%%%%%%%%%%%%%%%%%%%%%%%%%%%%%%%%%%%%%%%%%
%                       TITLE PAGE                                         %
%%%%%%%%%%%%%%%%%%%%%%%%%%%%%%%%%%%%%%%%%%%%%%%%%%%%%%%%%%%%%%%%%%%%%%%%%%%%
\Title{\vbox{\hbox{hep--th/9712071}\hbox{IASSNS--HEP--97/127,
PUPT-1727}}}
{\vbox{\centerline{New Perspectives on Yang-Mills Theories}
\vskip8pt\centerline{With Sixteen Supersymmetries}}}
\vskip 0.2in

\centerline{Ori J. Ganor$^{a,}$\footnote{$^1$} {origa@puhep1.princeton.edu} and
Savdeep Sethi$^{b,}$\footnote{$^2$} {sethi@sns.ias.edu} }
\vskip 0.1in
\medskip\centerline{\it$^a$Department of Physics, Princeton University,
Princeton, NJ 08544, USA}
%\vskip 0.1in
\medskip\centerline{\it $^b$School of Natural Sciences, Institute for Advanced
Study,
Princeton, NJ 08540, USA}

%%
%% Draft of paper defining compactified brane theories.
%%

\vskip 0.75in

\noindent

We describe various approaches that give matrix descriptions of compactified
NS five-branes. As a
result, we obtain matrix models for Yang-Mills theories with sixteen
supersymmetries in dimensions $2,3,4$ and $5$. The equivalence of the various
approaches relates the Coulomb branch of certain gauge theories to the moduli
space
of instantons on $\MT{4}$. We also obtain an equivalence between certain
six-dimensional
string theories. Further, we discuss how various perturbative
and non-perturbative features of these Yang-Mills theories appear in their
matrix
formulations. The matrix model for four-dimensional Yang-Mills is manifestly
S-dual.
In this case, we describe how
 electric fluxes, magnetic fluxes and the
interaction between vector particles are realized
in the matrix model.

%\vskip 0.08in
\Date{11/97}
%\draftmode
\baselineskip=20pt plus 2pt minus 2pt

% =====================================================================
%
% Introduction
% =====================================================================
%

\newsec{Introduction}

The matrix model formulation of M theory \rBFSS, holds the promise of
providing a non-perturbative description of M theory, and consequently
string theory. Perhaps of equal importance, this description of M theory
suggests that quantum field theory might be formulated in a new way. In
this new description, some symmetries, such as Lorentz invariance, may
no longer be manifest; yet other properties like S-duality may become manifest.
The aim of this paper is
to propose non-perturbative matrix formulations of Yang-Mills theories
in various dimensions quantized in the discrete light-cone formalism
(DLCQ) \rdlcq. We will consider theories with sixteen
supersymmetries,\foot{To orient the reader with our conventions, note that
$\SUSY{4}$
Yang-Mills in four-dimensions has sixteen real supersymmetries.}
although the main features of our construction seem to go over to cases
with less supersymmetry. The case we understand best is Yang-Mills
at a fixed point, and our discussion will focus on the matrix formulations of
these conformal field theories.

Let us start with a configuration of parallel M theory five-branes.
Recall that the theory on $k$ coincident M theory five-branes is the
$(2,0)$ field theory, first found in \refs{\rwtensor, \rstensor}. At low
energies, this theory flows to an interacting
superconformal fixed point in six dimensions. See \rreview\ for a
review. After wrapping the five-branes on a circle, we obtain a theory
which at low energies is five-dimensional Yang-Mills with a $U(k)$ gauge
group and minimal matter content. The square of the Yang-Mills coupling
constant, $g^2_{YM}$, has dimension $3-d$ for Yang-Mills in $d$ spatial
dimensions. Therefore in five-dimensions, the theory becomes free at
low-energies.

When the five-branes are compactified on a two-torus, we obtain a
natural description of four-dimensional Yang-Mills. This theory has a
Coulomb branch parametrized by the expectation values for the scalar
fields. The theory at the origin of the moduli space is an interacting
conformal field theory characterized by the dimensionless Yang-Mills
coupling constant. The theory is believed to be S-dual
\refs{\rmont, \rsdual}. In this construction, S-duality is made
manifest in terms of the symmetry group of the torus \rwtensor. A
matrix definition of the wrapped five-branes should therefore give a
formulation of
Yang-Mills with manifest S-duality.

Compactification on a three-torus gives three-dimensional Yang-Mills at
low-energies. The coupling constant is again dimensionful and we are
driven to a strong coupling fixed point in the infra-red. The Yang-Mills
theory has a manifest $Spin(7)$ symmetry acting on the seven scalars and
associated fermions. However, the superconformal fixed point actually
has $Spin(8)$ global symmetry, and is believed to be interacting
\refs{\rSS, \rBS, \rreview}. For the abelian case, the extra dimension
can be viewed as arising from the scalar dual to the gauge-field in
three-dimensions.

Lastly, compactification on a four-torus gives two-dimensional
Yang-Mills. In the infra-red, we are again driven to a strong coupling
fixed point. In this case, the resulting two-dimensional sigma model is
believed to have a target space given by the classical moduli space of
vacua \refs{\rBS,\rDVVstring}. For gauge group $U(k)$, the moduli space is
\eqn\moduli{ \CM = { (\IR^8)^k \over {\bf S}_k},}
and the infra-red theory is an orbifold conformal field theory. Note that
in this case, the spectrum for the two-dimensional theory
in DLCQ is discrete since there are no non-compact dimensions.

While the goal of this discussion is to provide matrix descriptions for
each of these theories, these Yang-Mills theories themselves define M
theory compactifications in the DLCQ formalism. We can interpret each of
the features described above in terms of some property required by
compactified M theory. The two-dimensional Yang-Mills theory describes M
theory on a circle. In the limit where the circle goes to zero size, the
Yang-Mills coupling is driven to infinity and so the orbifold conformal
field theory should describe the type IIA string with $k$ units of
longitudinal momentum \refs{\rmotl, \rBS, \rDVVstring}. Since the
infra-red theory is an orbifold conformal field theory, the resulting
Hilbert space agrees with the Hilbert space for a free string.  Our
matrix description of this theory can then be viewed as a matrix model
for the matrix model of the type IIA string. Likewise, the type IIB
string is defined in terms of the three-dimensional Yang-Mills theory,
where the enhanced flavor symmetry reflects the appearance of an extra
dimension needed for the ten-dimensional string theory. The existence of
S-duality in four-dimensional Yang-Mills is needed for the T-duality
symmetry which exists in M theory compactified on a three-torus
\refs{\rlenny, \rori}.

There are several different ways to obtain matrix formulations of Yang-Mills
theories. One approach is to determine the matrix description of wrapped M
theory
five-branes. In a suitable limit, this matrix theory will describe the
compactified
$(2,0)$ field theory. This is an interesting question which involves a study of
impurity theories similar to the system described in \rsIIB. A second approach
is
to start with a non-critical string theory in six
dimensions: either the $(2,0)$ string theory living on type IIA NS
five-branes or the $(1,1)$ string theory living on type IIB NS
five-branes \rsdecoupled. On compactification, these two theories are
part of a connected moduli space. The advantage of studying the string theory
rather
than the field theory is that there are two different matrix models for the
compactified
string theory. One is an impurity model which is quite difficult to analyze
while the
second is a more conventional gauge theory. There are related formulations
which can be
reached by a series of dualities from one of the above prescriptions, and we
will explain one additional formulation. When restricted
 to energies $E \ll M_s$, where $M_s$ is
the string scale, these models have branches describing Yang-Mills fixed
points.

In the following section, we describe how to obtain matrix models from
some of these various approaches and the relations between them. Along the way,
we
will describe a duality between certain six-dimensional decoupled theories. The
duality has the interesting feature of exchanging $N$ with a different
parameter and
may shed some light on the large $N$ limit of matrix theory.  We will
also find relations between the Coulomb branch of certain quiver gauge theories
and
degenerations of the moduli space of instantons on $\MT{4}$. In section three,
we
consider the conditions under which we can reduce to quantum mechanics and
obtain
a matrix formulation of the Yang-Mills fixed points. In section four, we
explore the
 properties of supersymmetric Yang-Mills (SYM) from the matrix model
description. We discuss electric and magnetic
fluxes, Wilson lines in the longitudinal direction, vector particles, Coulomb
interactions and R-symmetries. This includes a discussion of the perturbative
limit
of four-dimensional Yang-Mills where we show how
the force between vector bosons can be reproduced in the matrix model. This is
quite
non-trivial since in a description where S-duality is manifest, the
perturbative
 limit is not distinguished.  Finally in
 section five, we conclude by discussing generalizations to cases with less
supersymmetry, and by mentioning some of the many open questions. Some related
issues
have been discussed recently in \rBR.

% =====================================================================
%
% Section (1) : Various Approaches to Compactified String Theories
% =====================================================================
%

\newsec{Various Approaches to Compactified String Theories}

% --------------------------------------------------------------------- %
% Compactifying string theories on \MT{4}
% --------------------------------------------------------------------- %
\subsec{ Compactified six-dimensional string theories}

Four-dimensional $\SUSY{4}$ SYM can be viewed as a limiting case of a more
general theory obtained by compactifying a 5+1D field theory on a
torus, $T^2$. We could compactify either 5+1D SYM with
$(1,1)$ supersymmetry or the $(2,0)$ field theory living on parallel
five-branes.
In the second case, the limit where we recover 3+1D SYM requires taking $T^2$
to
zero size while holding fixed the complex structure $\tau$. The coupling in the
resulting SYM theory is $\tau$ and $SL(2, \IZ)$ is manifest. In turn, the
$(2,0)$
or $(1,1)$ field theories describe the low-energy excitations of complete
theories
living on, respectively, type IIA or IIB NS five-branes \rsdecoupled. We can
therefore
 start with the more general question of how to provide a matrix model for the
string theory compactified on ${\bf T}^d$ for $d=1,2,3,4$. The matrix model for
SYM in
D+1 dimensions can be deduced by taking a limit of the external parameters in
the
model for $d=5-D$.

Compactifying $k$ NS five-branes on $\MT{4}$ gives an effective theory with
sixteen
supersymmetries which we will call $\X_k$. For a generic torus, $\X_k$ is 1+1D
and contains
all of the KK states of the compactified six-dimensional string theory.
T-duality
of the compactified
string theory implies that $\X_k$ depends on an {\it external} parameter,
$$
u \in SO(4,4,\BZ) \backslash SO(4,4,\BR) / (SO(4)\times SO(4)),
$$
where $u$ parametrizes the shape and size of $\MT{4}$ and the
choice of $B_{\mu\nu}$.  The matrix model for the $\pV = N$
sector of $\X_k(u)$
is a theory $\X_{k,N}(u)$ with eight supersymmetries which
depends on the external parameter $u$.
Since $\X_k(u)$ has BPS strings, $\X_{k,N}(u)$ should generically be a 1+1D
conformal field theory compactified on $\MS{1}$. There are many different
 but related ways to define $\X_{k,N}(u)$.
We will discuss the following three ways of obtaining matrix formulations.
\vskip 0.1in

\item{Route 1:}
We can study the theory localized at the singularity of M theory on $T^d \times
A_{k-1}$. This geometric picture is particularly nice since it extends easily
to
the case of $D$ and $E$ singularities.
When $d>3$, this theory needs to be defined in terms of the 5+1D string
theories
discussed in \rkin. These string theories are obtained by placing N type IIB NS
five-branes near an $A_{k-1}$ singularity. The case of type IIA five-branes has
not been considered but we will demonstrate a duality relating type IIA to type
IIB. This duality is quite interesting since it exchanges $N$ and $k$.

\item{Route 2:}
We can consider the matrix definition of $k$ longitudinal type IIA NS
five-branes
wrapped on $\MT{4}$. The matrix model for this case is defined in terms
of the decoupled theory on $N$ NS five-branes wrapped on the dual torus,
 $ \widehat{T}^4 \times \widehat{S}^1$. We will
study the sector with $k$ units of string winding on $ \widehat{S}^1$. This
is the model obtained by probing wrapped NS five-branes in string theory.

\item{Route 3:}
Lastly, we can take the theory living on $k$ NS five-branes, wrapped on $\MT{4}
\times
 \MR{1,1}$, and view it as a compactification
of a  six-dimensional
Lorentz invariant theory decoupled from gravity and string theory \rsdecoupled.
A description
of  $\X_{k,N}(u)$ can then be obtained by applying the $SO(5,5,\IZ)$ T-duality
to the
prescription given in \refs{\rSenTn, \rSeiWHY} which relates the DLCQ theory to
the
theory on a small space-like circle.

% --------------------------------------------------------------------- %
% Route 1:  M-theory on an $A_{k-1}$ singularity
% --------------------------------------------------------------------- %
\subsec{Route 1:  From M-theory on an $A_{k-1}$ singularity}

The matrix model for M theory on $ T^d \times {\rm ALE}$ follows readily
from the results in \rdoumoo. The radii of the torus are  $ \{ R_1, \ldots, R_d
\}$.
 Let us specialize to the case of $T^d \times A_{k-1}$ for the moment. The
 theory is $d+1$-dimensional Yang-Mills with gauge
group,
\eqn\firstgauge{ U(N)_{1} \times \ldots \times U(N)_{k},}
and eight supersymmetries. The hypermultiplet content is encoded in the
extended
Dynkin diagram. Each link gives a hypermultiplet in the representation,
$ \oplus_{ij}\, a_{ij} (N_i, {\bar N}_j)$, where $a_{ij}$ is one for a
link between the $i^{th}$ and $j^{th}$ node and zero otherwise. For $A_{k-1}$,
this gives
hypermultiplets in the $(N, {\bar N})$ of $U(N)_i \times U(N)_{i+1}$ for
$i=1\ldots k$, where $k+1 \equiv 1$.  The gauge group and matter content will
actually change for
 $d \geq 3$ in a way which we will describe later. For
$d>3$, the gauge theory is no longer well-defined in the ultra-violet.
The natural definition is in terms of the decoupled six-dimensional
theory living on type II five-branes on the $A_{k-1}$ singularity \rkin.
When all the Fayet-Iliopoulos parameters are set to zero, there is a
Coulomb branch as well as a Higgs branch. The Coulomb branch describes the
physics
 localized at the singularity, which is a $7-d$-dimensional gauge theory.
 The Higgs branch should describe the spacetime physics which is clearly not
localized at the
 singularity. Our interest is primarily with the Coulomb branch in this
discussion. The
gauge theory associated to the singularity in M theory has a
coupling constant,
\eqn\yangmillscoupling{ (\hat{g}_d)^2 = {1\over M_{pl}^{3} R_1 \ldots R_d }.}
The $D$ and $E$ cases are similar, differing only in the choice of gauge group
and
matter content.

What is the relation of M theory on $T^d \times A_{k-1}$ to compactified
six-dimensional string theories?   If $R_1$ is small, we can reduce from M
theory to
type IIA compactified on $T^{d-1} \times A_{k-1}$ with,
\eqn\scales{ M_s^2 = M_{pl}^3 R_1.}
The ten-dimensional type IIA string coupling is $g_s^A = M_s R_1$. When $R_2$
becomes small,
we should T-dualize to type IIB with string coupling $g_s^B = R_1/R_2$.
Describing
the compactified string theory via M theory on an orbifold space can be
directly related
to compactified NS five-branes by a T-duality argument. Let us start with $k$
wrapped type IIB
NS five-branes at a point on $\IR^3 \times \MS{1}$ where the radius of the
circle is $T$.
T-dualizing on $S^1$ takes the NS five-branes to type IIA Kaluza-Klein
monopoles with the
scale of the metric set by $1/M_s^2 T$. In the limit where $T$
becomes small, the Kaluza-Klein monopole degenerates to an $A_{k-1}$
singularity
\refs{\rmultitn, \rashoke}.  In this limit, we can use the matrix theory
described in \firstgauge.

The $k$ NS five-branes are then at a point on $\IR^3 \times \MS{1}$ where the
circle is
very small. However, from the perspective of the gauge
theory on the brane, the circle actually has radius $ 1/R_1$ \rnati. We can
make this
 circle large by taking $R_1$ small. The limit that we will want to study
results in a
theory parametrized by $T^{d-1}$ and $M_s$, and is obtained by taking:
\eqn\limits{ \eqalign{  M_{pl} \r \infty, \qquad R_1 \r 0, \qquad T \r 0. \cr
}}
In this region of parameter space, the theory localized at the singularity will
describe
the compactified six-dimensional string theory. With $M_s$ held fixed, it will
interpolate between the compactified $(1,1)$ and $(2,0)$ string theories
as we vary the parameters of $T^{d-1}$. Finally,
considering energies much smaller than $M_s$ gives the infra-red behavior of
Yang-Mills
compactified on $T^{d-1}$. We can therefore conclude that by studying the
Coulomb branch
of matrix theory on $T^d \times {\rm A_{k-1}}$, we will obtain matrix
definitions of
Yang-Mills in various dimensions.

\subsec{Interpolating between six-dimensional string theories}

The matrix theory torus is `dual' to the M theory torus. In terms of the
longitudinal direction with size $\RV$, the torus has radii:

\eqn\sigrr{\eqalign{
\Sigma_1 &= {1\over {M_{pl}^3 \RV R_1}}={1\over {M_s^2 \RV}},
\cr
\Sigma_j &=  {1\over {M_{pl}^3 \RV R_j}}= {{R_1}\over {M_s^2 \RV R_j}},
\qquad j=2\dots d.
\cr
}}

The case $d=1$ corresponds to the decoupled theory on type IIB NS
five-branes \refs{\rsIIB, \reIIB}. In this case, the Coulomb and Higgs
 branches are expected
to decouple in the infra-red \rwtensor. As a first approximation to the
physics of the Coulomb branch, we can use a moduli space approximation
and describe the low-energy dynamics by a sigma model with a metric that
has a `tube' structure. This is unlikely to be a
good description of the physics at short distances. The failure of the
moduli space approximation already occurs for M theory on an $A_{k-1}$
singularity where $d=0$. Quantum mechanics on the Coulomb branch can be
approximated
by a sigma model with a metric that behaves as $1/r^3$ at short
distances in five-dimensions. However, the gauge bosons should appear as
$L^2$ ground states in the quantum mechanics but these states
generally cannot be seen without including more degrees of freedom. It
seems likely that the Coulomb branch theory in $1+1$-dimensions also
requires more degrees of freedom to be sensible.

Let us turn to $d=2$ keeping $M_s$ finite. In this and subsequent cases,
 there is no issue about the small distance behavior of the Coulomb branch
metric since
 there is a genuine  moduli space.   How do we see
T-duality? As $R_1 \r 0$, $\S_2 \r 0$ so the theory becomes
$1+1$-dimensional. The matrix model coupling is,
\eqn\twodim{ g^2 = {\RV \over R_1 R_2}.}
The dynamics in this theory is governed by the dimensionless parameter,
\eqn\param{ \g = g^2 \S_2 = {1\over (R_2 M_s)^2}.}
When $R_2 \r \infty$, $\g \ll 1$ so the effective gauge interactions are
weak. We can then dimensionally reduce to two-dimensions and flow to the
Coulomb branch conformal field theory. The Coulomb branch is
parametrized by $4Nk$ scalars and describes the decoupled theory on type
IIB NS five-branes.

  When $R_2 \r 0$, $ \g \gg 1$ so we first flow to the three-dimensional
fixed point. Including the dual scalars, the Coulomb branch is again
parametrized by $4Nk$ scalars, as we expect for a six-dimensional
theory. The  new direction emerges as the periods for the dual scalars
decompactify. The Coulomb branch is quantum corrected but we can use a
three-dimensional mirror to determine the metric
\refs{\rmirror,\rconjecture}. Including quantum corrections, the moduli space
corresponds to the moduli space of $N$ instantons in $SU(k)$
gauge theory. Reduction to two-dimensions then gives the matrix
description of the decoupled theory on type IIA NS five-branes. This
matrix description can also be obtained from the Higgs branch of the
two-dimensional model with $U(N)$ gauge symmetry, an adjoint
hypermultiplet and $k$ fundamentals \refs{\rquantumfive, \rstringfive}.
 This
two-dimensional analogue of mirror
symmetry is the simpler side of the duality described in
\rsIIB, since it follows immediately from three-dimensional mirror
symmetry. Note that in both limits, we see a six-dimensional
theory on the Coulomb branch. In both the type IIA and type
 IIB cases, knowing
the moduli space is unlikely to be sufficient to describe the IR limits
of the Higgs or Coulomb branch theories. Without a better understanding of
the IR limits, the matrix definition is still largely implicit.

% --------------------------------------------------------------------- %
% A duality between decoupled theories
% --------------------------------------------------------------------- %
\subsec{A duality between decoupled theories}

At this point, we cannot resist explaining an interesting duality between
 six-dimensional
string theories. This duality was conjectured by Intriligator 
\rkin.\foot{We wish to thank K. Intriligator for bringing this to our
attention.} As we 
just mentioned, matrix models for SYM can be obtained by
 compactifying
the six-dimensional string theories associated to type IIB five-branes at $ADE$
 singularities.
Consistency conditions for these theories have been described in \rjulie. We
 could ask similar
questions about type IIA five-branes at $ADE$ singularities. On
 compactification, the theory
 of $N$ type IIA five-branes at an $ADE$ singularity is related by T-duality
 on a longitudinal
 circle to the IIB case but in six-dimensions, the theories are distinct
 and we obtain no
equivalence this way. Let us take $N$ IIB
 NS five-branes
on an $A_{k-1}$ singularity. We can equivalently consider $N$ type IIB
 five-branes on a
KK monopole
with an $A_{k-1}$ singularity. In the decoupling limit, as discussed above,
 the period for
the scalar parametrizing the compact transverse circle decompactifies. We are
 then free to
T-dualize on the compact direction.

Under this T-duality, the $N$ IIB five-branes turn into a type IIA KK monopole
 with an
$A_{N-1}$ singularity. The type IIB KK monopole turns into $k$ type IIA NS
 five-branes. With
no additional compact directions, we are free to take the decoupling limit
 in both pictures.
This gives an equivalence between the decoupled theory on $N$ IIB five-branes
 at an $A_{k-1}$
singularity and the decoupled theory on $k$ type IIA five-branes at an
 $A_{N-1}$ singularity.\foot{This exchange of $N$ and $k$ was noted
 independently in a geometrical construction of
four-dimensional gauge theories \rKMV.}
In particular, this duality allows us to trade the usual large $N$ limit of
 the matrix model,
about which we know little, for a study of the matrix model with fixed $P^{+}$
 but
on a highly singular space. We might hope to gain some understanding of the
 large $N$ limit
from this approach and this is currently under investigation.

By including orientifold actions, this argument generalizes to the case of $D$
 singularities
since we can replace a $D$ singularity by an ALF space analogous to the
 multi-Taub-NUT
metric associated to $A$ singularities.
We can also generalize it to the case of type IIB $(p,q)$ five-branes at
 an $A_{k-1}$
singularity by considering an M theory dual without a conventional type IIA
 description. The
 dual of a $(p,q)$ five-brane described in \reIIB\
is M theory on $X_{p,q} = (\IC^2 \times \MS{1})/ \IZ_q$. In our case, the
 dual simply becomes
$k$ M theory five-branes on $X_{p,q}$. Since this leads us away from our main
discussion, we will not explore further these theories and their
 generalizations here.

% --------------------------------------------------------------------- %
% Route 2:  Longitudinal five-branes in M theory
% --------------------------------------------------------------------- %
\subsec{Route 2:  Longitudinal five-branes in M theory}

We can obtain another definition for $\X_{N,k}(u)$
and its low-energy limit describing the $(2,0)$ field theory on $\MT{4}$
by compactifying $k$ parallel longitudinal type IIA or M theory five-branes.
We will see that this route leads us
to study the low-energy descriptions of gauge theories with point-like or
 string-like impurities in dimensions greater than four. For related comments
see \rextraDVV.
As explained in \rsIIB, interesting 1+1D theories can be obtained
by inserting extra localized 1+1D degrees of freedom into a 2+1D theory
with local gauge fields.
The relevant question for us is how
to elevate the impurity construction to cases where the bulk gauge theory is
five or six-dimensional.
We will define these bulk theories in terms of the decoupled theory on parallel
NS five-branes wrapped on $\MS{1} \times \MHT{4}$ \rsdecoupled.
We will argue that in these cases, the impurities become dynamical in the sense
that they can be constructed from {\it internal} degrees of freedom
of the system. The dynamical impurity is in essence a generalization of the
brane-probe
technique. We apply this approach to 0-brane probes of
the $(2,0)$ 5+1D CFT and 1-brane probes of the $(2,0)$ string theory.

% - - - - - - - - - - - - - - - - - - - - - - - - - - - - - - - - - - - %
% Compactifying the B-D model
% - - - - - - - - - - - - - - - - - - - - - - - - - - - - - - - - - - - %

We can obtain a matrix definition of five-branes wrapped on a torus, ${\bf
T}^d$ with $d \leq 4$, by starting with the matrix model for a longitudinal
five-brane \rbd\ extended to $k$ five-branes. This model contains a $U(N)$
vector multiplet $X$, an adjoint hypermultiplet $H$ together with $k$
hypermultiplets, $Q^f$, in the fundamental representation. The flavor index
runs from $1$ to $k$. Let the worldvolume of the four-branes fill $x_1, \ldots,
x_4$. The location of the four-branes is then specified by their positions in
$x_5, \ldots, x_9$. We will only be concerned with the case where the
four-branes are very close together, or actually coincident. It is interesting
to note that the effective $U(k)$ gauge group appears as a {\it flavor}
symmetry in the matrix description. This is particularly interesting when we
consider large $k$ color expansions.

The four scalars of the hypermultiplet $H$, denoted $H^a$ for $a=1\dots 4$,
parameterize
the position parallel to the longitudinal five-branes.
We can derive the matrix model for 3+1D SYM
from a limit of the $(2,0)$ theory
compactified on $\MT{4}$ and the latter can be obtained from
the low-energy limit of $k$ parallel M theory five-branes \rstensor.
Specifically, we take the limit where the distances between pairs of
five-branes, $x_0^{(i)} - x_0^{(j)}$, goes to zero and where we restrict to
energies
of order $E\sim M_{pl}^{3/2} |x_0^{(i)} - x_0^{(j)}|^{1/2}$.
In this way $E^2$ is of the order of the tension of BPS
strings in the $(2,0)$ theory.
The six-dimensional conformal field theory  can be studied by taking the limit
where the
five-branes are coincident i.e. by setting all the positions $x_0^{(i)}$ to
zero.

To find the matrix description for five-branes wrapped on a circle with radius
$R_1$, we
take the original system as well as all its translates along $x_1$. The gauge
group becomes infinite-dimensional and we then quotient out by the symmetry
group generating discrete translations along $x_1$ by multiples of $2\pi R_1$
\refs{\rwati, \rBFSS, \rori}. In practice, we take a large gauge
transformation, $\Omega \in U(\infty)$, and restrict to matrices obeying,

\eqn\modout{\eqalign{
H_1 = \Omega^{-1} H_1 \Omega + 2\pi R_1, &\qquad
H_a = \Omega^{-1} H_a \Omega,\qquad a=2\dots 4,\cr
Q^f = {\Lambda_{f'}}^f\Omega^{-1} Q^{f'}, &\qquad
X_\u = \Omega^{-1} X_\u \Omega,\qquad \u=5\dots 9,\cr
}}
where $\Lambda$ is some fixed $k\times k$ flavor matrix. This matrix encodes
the choice of Wilson line along $x_1$ for the gauge-field on the parallel
four-branes. These constraints are solved by taking $X_1= -i \partial_1 - A_1$,
and by promoting the rest of the $X$ and $H$ variables to fields depending on
the periodic coodinate $x_1$. In this representation, $ \Omega$ is diagonal
with eigenvalues $e^{ix_1/R_1}$. However, the constraint on the hypermultiplets
is solved by localizing the hypermultiplets to $k$ specific points $x_1 =
\lambda_i$ where $i=1,\ldots, k$ and where $\lambda_i$ is an eigenvalue of the
matrix $\Lambda$. Thus the system becomes a bulk 1+1D SYM with sixteen
supersymmetries and with $k$ impurities at fixed positions $\lam_i$ (see also
\rsIIB). The impurities explicitly break half the supersymmetries.

A similar argument applies when the five-branes are wrapped on $\MT{d}$. The
bulk physics is then described by a $d+1$-dimensional $U(N)$ gauge theory with
a $d+1$-dimensional adjoint hypermultiplet. The hypermultiplets in the
fundamental representation are quantum mechanical and live at specific points
on $\MHT{d}$. These hypermultiplets treat the spatial components of the
connection, $A_{\mu}$, as scalars. In particular, the coupling of the
hypermultiplets to the fields $A_1,\ldots, A_d$ breaks Lorentz invariance.
By choosing appropriate limits for the size of the compact directions, we
should be able to recover a complete description of longitudinal $p$-branes
with $p=1,2,3$ and $4$. This description should capture the spacetime metric as
well as the physics localized on the brane itself. In this discussion, we are
only concerned with the theory on the five-branes.

% - - - - - - - - - - - - - - - - - - - - - - - - - - - - - - - - - - - %
% 4+1D
% - - - - - - - - - - - - - - - - - - - - - - - - - - - - - - - - - - - %

When we further compactify on $\MT{4}$ we naively find
4+1D SYM with 0+1D impurities.
According to \rfourtorus\ we know that 4+1D SYM should
be interpreted as the $(2,0)$ 5+1D theory compactified
on $\MS{1} \times \MHT{4}$.
How should the impurity be interpreted?

We claim that the impurity is no longer external but is defined
as the sector with $k$ units of KK momentum along $\MS{1}$.
In this way we also make contact with the description of
longitudinal objects as momentum states in the matrix model (see \rImamura).
To justify the claim that at low energies $k$ KK states look
point-like we can estimate the various energy scales involved.
We are looking for an appropriate description
of the states at energy levels:
$$
E = {{2\pi k}\over r} + \epsilon,\qquad \epsilon \ll {1\over r},
$$
where $r$ is the radius of the circle $\MS{1}$.
% - - - - - - - - - - - - - - - - - - - - - - - - - - - - - - - - - - - %
% The bulk
% - - - - - - - - - - - - - - - - - - - - - - - - - - - - - - - - - - - %
The low-energy theory in 4+1D without the momentum has a moduli
space of $(\IR^5)^k/S_k$. Even at the origin of the moduli space the
theory is free in the infra-red.
At a generic point in the moduli space with a vacuum expectation value (VEV),
$v = \sqrt{\sum_1^5|\phi_i|^2}$, there are $W$ bosons of mass
$r^{1/2} v$ and also monopoles which are strings with tension
$r^{-1/2} v$. The energy scale set by the strings is higher
by a factor of $r^{-3/4} v^{-1/2}$.
As long as $v\ll r^{-1}$ both scales are much smaller than the
compactification scale.
States with energies which satisfy,
\eqn\esat{
E\ll  r^{-1},
}
can be described by the effective (non-renormalizable) Lagrangian
of 4+1D SYM with dimensionful coupling constant,
$$
{1\over {{  g_{\rm YM}^2}}} = r^{-1}.
$$
% - - - - - - - - - - - - - - - - - - - - - - - - - - - - - - - - - - - %
% Adding momentum
% - - - - - - - - - - - - - - - - - - - - - - - - - - - - - - - - - - - %
Let us add
one unit of momentum along the small $\MS{1}$, and ask what it
looks like in the low-energy limit in 4+1D.
The momentum state becomes a heavy soliton which we locate at the
origin. It breaks half the supersymmetries.
The original symmetries were the $SO(4,1)_N$ Lorenz invariance
and $Sp(2)_R$ R-symmetry.
The soliton breaks $SO(4,1)_N$ down to $SO(4)_N$ while $Sp(2)_R$
remains intact. We can realize this as a D0-brane on a D4-brane, making
$Sp(2)\sim SO(5)$ geometrical.
The unbroken supersymmetries transform as (see also \rquantumfive):
$$
(\rep{2},\rep{1})_N\otimes\rep{4}_R
$$
%%%(recall that both $\rep{2}$ of $SU(2)$
%%% and $\rep{4}$ of $SO(5)$ are pseudo-real)
with a reality condition.
An anti-0-brane will transform as
$$
(\rep{1},\rep{2})_N\otimes\rep{4}_R.
$$
At a generic point in moduli space,
the soliton is described by an effective quantum mechanics.
The effective quantum mechanics contains four bosons $X^i$ and
eight fermions.
The bosons are in the $(\rep{2},\rep{2})_N\otimes \rep{1}_R$
while the fermions are in the $(\rep{1},\rep{2})_N\otimes\rep{4}_R$
with a reality condition.

% - - - - - - - - - - - - - - - - - - - - - - - - - - - - - - - - - - - %
% Instanton moduli space
% - - - - - - - - - - - - - - - - - - - - - - - - - - - - - - - - - - - %
At the origin of moduli space the KK state behaves like an
instanton of 4+1D SYM.
The low-lying states which are related to the impurity can
be treated by quantizing the instanton moduli space.
Thus the 0+1D theory
has four more modes -- the size and orientation -- which parameterize
the moduli space $\IR^{4}/\IZ_2$.

When the instanton has a finite size, it is no longer obvious
that the low-energy description in terms of a 0+1D quantum mechanics
interacting with a 4+1D bulk is still adequate.
We can justify it in the following way.
Suppose we take a small UV cutoff $\epsilon$ on the energies.
The characteristic length scales of the IR processes we
are describing will be $\epsilon^{-1}$. To justify the 0+1D
picture, we have to show that we can make a small wave-packet
in instanton moduli space that will localize on instanton sizes
$\rho\ll\epsilon^{-1}$; yet the energy spread of the wave-packet
must be much smaller than $\epsilon$.
The Hamiltonian for the collective modes coordinatizing instanton moduli
space is approximately,
$$
H \sim r^{-1}\{\dot{\rho}^2  + \rho^2 \tr (\Omega^{-1}\dot{\Omega})^2\}.
$$
where $\rho$ is the size of the instanton and $\Omega \in SU(2)/\IZ_2$ is the
orientation.
The low energy levels where $\epsilon$ is above $1/r$
correspond to wave functions $\psi$ which behave like
$$
\psi \sim e^{i(\epsilon/r)^{1/2}\rho}.
$$
Thus the typical scale for $\rho$ is,
$$
\rho\sim \epsilon^{-1/2} r^{1/2},
$$
and indeed, the bulk excitations with energy $\epsilon$ will have
an approximate size of $\epsilon^{-1}\gg \rho$.
This means that at energies $\epsilon\ll r^{-1}$, it is safe to assume
that the instanton is point-like.
% - - - - - - - - - - - - - - - - - - - - - - - - - - - - - - - - - - - %
% More on scales
% - - - - - - - - - - - - - - - - - - - - - - - - - - - - - - - - - - - %

After compactification on $\MT{4}$ the previous argument can no
longer be applied because the instanton is no longer localized.
However, we will show that the energy scale of the instanton
moduli space is much below the energy scale of momentum excitations.

Let us  compactify
the $(2,0)$ theory on $\MS{1}\times\MHT{4}$
with $\MS{1}$ of radius $r$ as before
and with $\MHT{4}$ of radii $\Sigma_1,\dots, \Sigma_4$
and take the limit $r\ll \Sigma_i$.
The energy scale of the instantons themselves is,
$$
M_{{\rm instantons}} \sim {1\over {g^2}} \sim r^{-1}.
$$
The other momentum states are at mass scale
$$
M_{{\rm momentum}} \sim \Sigma_i^{-1} \ll r^{-1}.
$$
The energy of an electric flux in the direction of $\Sigma_1$ is given
by
$$
M_{{\rm electric}} \sim
  {{r \Sigma_1}\over {\Sigma_2 \Sigma_3 \Sigma_4}} \ll \Sigma_1^{-1},
$$
and a magnetic flux in directions $\Sigma_1, \Sigma_2$ has energy
$$
M_{{\rm magnetic}} \sim
   {{\Sigma_3 \Sigma_4}\over {r \Sigma_1 \Sigma_2}} \sim r^{-1}.
$$
In the regime $r\ll \Sigma_i$, the low-energy is dominated
by the 0+1D quantum mechanics and the electric fluxes.
We can see that the energies coming from the 0+1D quantum mechanics
are of the same order of magnitude as the electric fluxes.
We can check this for the zero modes $X^i$ which represent the
center of mass of the instanton.
The momentum of the instanton is quantized in units of $1/\S_i$.
Since its mass is $r^{-1}$, the kinetic energy
will be of the order of $r / \Sigma_i^2$.
This is to be expected since the electric fluxes are obtained by
quantizing the global Wilson lines along $\MHT{4}$ which are part
of the instanton moduli space.

We can then conclude that the  matrix model for
the $(2,0)$ theory compactified on $\MT{4}$ is given by
quantum mechanics on the moduli space of $U(N)$ instantons with
instanton number $k$ on $\MHT{4}$.

% - - - - - - - - - - - - - - - - - - - - - - - - - - - - - - - - - - - %
% 5+1D
% - - - - - - - - - - - - - - - - - - - - - - - - - - - - - - - - - - - %

\subsec{Longitudinal five-branes in string theory}

Starting with the previous construction of $k$ M-theory five-branes wrapped
on $\MT{4}$, we can further compactify a direction transverse to the
five-branes. This way we obtain $k$ type IIA NS five-branes wrapped
on $\MT{4}$. The matrix formulation looks naively like 5+1D SYM compactified
on $\MHT{5}$ with $k$ 1+1D lines of impurities located parallel to one side of
$\MT{5}$.
Let us take the sides of $\MHT{5}$ to be $\Sigma_1,\dots,\Sigma_5$
and let the $k$ impurities be parallel to $\Sigma_5$.
The bulk 5+1D SYM has to be defined using the theory on $N$ type IIB
five-branes
with the $SO(5,5,\BZ)$ T-duality group \rsdecoupled. This theory has
string-like BPS excitations and so we interpret the 5+1D
SYM with impurities as the sector with $k$ units of winding number
along $\Sigma_5$. The impurities are dynamical as in the previous case.
Applying $SO(5,5,\BZ)$ T-duality, we can map the sector with $k$
units of winding number to a sector with $k$ units of momentum
along a dual $\widetilde{\Sigma}_5$. In this way, we again make contact
with the identification of longitudinal objects and KK momentum
in the matrix model.

Now let us take the limit $\Sigma_5\rightarrow\infty$, keeping
$\Sigma_1,\dots,\Sigma_4$ fixed and look for a low-energy description
of the resulting 1+1D theory.
To be more generic we can put the string theory
on $\MT{4}\times\MR{1,1}$ with $k$ strings stretched along the
uncompactified direction. We denote this 1+1D theory by
$\X'_{k,N}(u)$.
It depends via $u$ on the sixteen external parameters which parameterize,
$$
 {\cal M}' = SO(4,4,\BZ)\backslash SO(4,4,\BR) / (SO(4)\times SO(4)).
$$
In the spacetime interpretation of the matrix model, the limit
$\Sigma_5\rightarrow
\infty$ corresponds to taking the type IIA string
coupling $g_s\rightarrow 0$. Thus $\X'_{k,N}$ is a matrix model
for $k$ wrapped NS5-branes in type IIA at $g_s=0$ compactified on $\MHT{4}$.

% --------------------------------------------------------------------- %
% Route 3:  From Lorentz invariance
% --------------------------------------------------------------------- %
\subsec{Route 3:  From Lorentz invariance}

We could also derive the 1+1D theory $\X_{k,N}(u)$ by a straightforward
application of the rules of \rSeiWHY.
To find the sector $\pV = N$ of
$\X_k(u)$ in the DLCQ, we can consider the string theory on $k$ five-branes
 with parameters,
\eqn\newparam{ \widetilde{M_s}, \widetilde{R_i}, R_s }
and $N$ units of KK momentum along $R_s$.
This theory is related to the $\X_{k,N}(u)$ by a Lorentz transformation. We can
fix the parameters \newparam\ in terms of $M_s$, the original transverse
 dimensions $R_i$
and the longitudinal direction $\RV$:
$$ R_s \widetilde{M_s}^2  = M_s^2 \RV. $$
As $R_s \r 0$, the radii $ \widetilde{R_i} \r 0$ and $ \widetilde{M_s} \r
 \infty$.
Using the $SO(5,5, \IZ)$ T-duality, we can exchange the shrinking circle $R_s$
 for a circle
with finite radius, $ 1/M_s^2 \RV$. The momentum is exchanged for $N$ units of
 string
winding. This leads to a matrix formulation in terms of $k$ NS five-branes
 wrapped on $\MT{4}$
with sides $ \widetilde{R_i}$. The parameters for this torus are again
 determined by the choice
of $u$. We have performed a single T-duality to arrive at this description,
 but $\X_{k,N}(u)$
has two low-energy limits corresponding to the theory on either type IIA or
 type IIB
five-branes. It is then convenient to T-dualize again on one of the transverse
 circles which
sends,
$$ u \,\,\, \r \,\,\, T(u), $$
where the map $T$ is independent of our choice of transverse circle. We can
 define the map
by picking an element,
$ t \in O(4, \IZ) \times O(4, \IZ), $
with $\det(t) = -1$. Then $T(u) = \lbr t \circ g \circ t^{-1} \rbr$ where
 $g\in SO(4,4,\IR)$
is any representative for $u$. This construction is independent of the choice
 of $t$. We will
need this map to relate this proposal to the previous ones.

% --------------------------------------------------------------------- %
% Relations between the various approaches
% --------------------------------------------------------------------- %
\subsec{Relations between the various approaches}

We have obtained three seemingly different matrix formulations for the
 compactified
six-dimensional string theory. How are they related? Let us start by
 considering the
matrix model describing energies $E \ll M_s$ which corresponds to
the $(2,0)$ field theory on $\MT{4}$. We have seen that
route (2) leads to quantum mechanics on the moduli space
of $U(N)$ instantons on the dual $\MHT{4}$ at instanton number $k$, while
route (3) leads to $U(k)$ instantons on the original $\MT{4}$ at
instanton number $N$. In this case, the equivalence follows from the
 Fourier-Mukai
transform \rmukai, or from T-duality of a system of $N$ 4-branes and $k$
 0-branes
on $\MT{4}$.

In fact, we have a more general statement that $\X_{k,N} (u)$ is equivalent to
$\X_{N,k}(T(u))$ which implies the above relation at low-energies. This
 equivalence
follows from the T-duality argument of section {\it 2.3} and proves that
 route (2) and
route (3) give equivalent formulations. By contrast, route (1) leads to a
 matrix description
 in terms of the Coulomb branch for the Yang-Mills theories \firstgauge\ in
various dimensions.  The Coulomb branch for the theory
of $N$ NS five-branes wrapped on $T^3$ at an $A_{k-1}$ singularity is a
hyper\kh
manifold. This compactified string theory is parametrized by
$$ SO(3,3,\BZ) \backslash SO(3, 3,\BR) / (SO(3)\times SO(3)),$$
which is
$$ \sim  SL(4,\BZ) \backslash SL(4, \BR) / SO(4).$$
So the moduli space is parametrized by the shape of a $\MT{4}$.
The equivalence of this approach with routes (2) and (3) implies that this
is the moduli space of $U(k)$ instantons on $\MT{4}$ with instanton number
 $N$.\foot{That the Coulomb branch is given by the moduli space of instantons
 on $T^4$
was suggested by probe arguments in \rkin.}
 In the
next section, we will see that a particular degeneration of this hyper\kh space
 does
indeed agree with the solution of
3+1D $\SUSY{2}$ quiver gauge theories found by other approaches
 \refs{\rWitBR, \rKMV}.\foot{Some related observations
 about theories with compact moduli spaces have
been made independently by \rGivKut.}

% --------------------------------------------------------------------- %
% (p,q) 5-branes
% --------------------------------------------------------------------- %
\subsec{$(p,q)$ Five-brane theories}

\def\MHS#1{{\widehat{{\bf S}}^{#1}}}

We can similarly study the matrix model for the $(p,q)$
five-brane theory compactified on $\MT{d}$. We will again go
along the three routes described above. We will see that
in each route, we can identify a certain $\BZ_q$ global symmetry
of $\X_{q,N}$ and twisting by the symmetry defines the $(p,q)$ five-brane
theory.
Let us first discuss the low-energy limit in $(6-d)$ dimensions.

% - - - - - - - - - - - - - - - - - - - - - - - - - - - - - - - - - - - %
% Low-energy limits
% - - - - - - - - - - - - - - - - - - - - - - - - - - - - - - - - - - - %
For a generic $\MT{d}$, the theory has $(6-d)$ uncompactified
dimensions and depends on $d^2$ external parameters,
$$
u\in SO(d,d,\BZ)\backslash SO(d,d,\BR) / (SO(d)\times SO(d)),
$$
which are the shape, size, and $B$-fields of $\MT{d}$.
Let
$$
r = {\rm gcd} (p,q).
$$
As explained in \reIIB, in 5+1D the low-energy limit is given
by $U(r)$ SYM, which at a generic point in the moduli space is
described by $r$ vector multiplets.
This is also true after compactification. In particular for
$d=2$, we have a 3+1D theory and the low-energy modes are generically
$r$ free vector multiplets. At the origin of moduli space,
the low-energy theory is the interacting theory to which $U(r)$
SYM flows in the IR, with a coupling constant given by
the area of $\MT{2}$ times $M_s^2$. This coupling constant
is defined up to S-duality.
Thus for generic $\MT{d}$, theories with different $(p,q)$ but
equal $r$ flow to the same IR theory. This is similar to the
situation in 5+1D where $p$ and $q$ entered only through
an irrelevant operator of dimension $6$, i.e. ${p\over q}\Tr F^3$.

For special values of $u$ the story is different.
Just like the NS5-brane theory, the $(p,q)$ theory has several
special limits for $u$ where a maximal number of dimensions
decompactify making the theory
six-dimensional.
 One such limit is when $\MT{d}$ becomes large and we are
back to the uncompactified $(p,q)$ five-brane theory.
Other limits are obtained when $\MT{d}$ is in the form
$\MT{d-d'}\times\MT{d'}$ with $\MT{d'}$ small and $\MT{d-d'}$ large.
To find the low-energy description in these limits,
we have to study how the $(p,q)$ theory behaves under T-duality.
This is easy to analyze using the realization of the theory in terms
of bound states given in \rsdecoupled. In this way, we can identify $p$ with
a D5-brane charge. After T-duality, we then get a system of NS five-branes
with some D-brane charges. The results are as follows:
for $d=1$, we start with the $(p,q)$ five-brane compactified on a very
small $\MS{1}$. This can be dualized to $q$ type IIA NS5-branes
on a large $\MHS{1}$. At low-energies, this theory contains $q$ tensor
multiplets with compact scalars,
and the $p$ charge is observed at low-energy energies as
a non-zero gradient for the compact scalars.
For $d=2$, we start with the $(p,q)$ five-brane compactified on a very
small $\MT{2}$. This can be dualized to the $(0,q)$ five-brane
on a very large $\MHT{2}$. The $p$ charge becomes magnetic flux
along this $\MHT{2}$. Note that this is consistent with the fact
that $p$ takes values in $\BZ/q\BZ$, i.e. it is defined modulo $q$.
A magnetic flux which is a multiple of $q$ can be embedded entirely
within the overall $U(1)$ factor of $U(q)$ and decouples from $SU(q)$.
Similarly, for $d=4$ we find the $(0,q)$ five-brane on a large $\MHT{4}$
with $p$ units of electric flux transverse to $\MHT{4}$.
For $d=3$ we find type IIA five-branes with some form of partial tensor
flux. This suggests that the $(2,0)$ theory also has a tensor
flux in $\BZ/q\BZ$.

% - - - - - - - - - - - - - - - - - - - - - - - - - - - - - - - - - - - %
% discontinous nature in p/q
% - - - - - - - - - - - - - - - - - - - - - - - - - - - - - - - - - - - %
As a consequence of this discussion, we can argue that the spectrum of
the theories is
not continous in the parameter $\theta = 2\pi {p\over q}$.
Recall that the coefficient
 of the term $\trp{F^3}$ in the derivative expansion of the
low-energy description of the $(p,q)$ theory is $\theta$ as argued in \reIIB.
The question then arose about whether the full $(p,q)$ theory is continuous in
$\theta$. See also the discussion in \rBarak.  The discussion above
 suggests that this is not the case, at
least for the compactified theories.
The Hamiltonian of the $(p,q)$ theory on $\MT{5}$ depends on
25 parameters which parameterize $\MT{5}$ and the $B$-fields.
The uncompactified theory is reached in a certain limit of these
parameters. As we have seen,
there are other low-energy limits obtained by
shrinking some sides of the $\MT{5}$ to zero and using T-duality.
We can use an element of $SO(5,5,\BZ)$ which maps the $p$ flux
into a magnetic flux and in this way, we reach another low-energy
limit which has the full $U(q)$ SYM gauge group.
Thus given a 5+1D theory, we might define the integer $q$ as
the maximal rank of the gauge groups attained in any of the
possible low-energy limits of the theory.

% - - - - - - - - - - - - - - - - - - - - - - - - - - - - - - - - - - - %
% Route (1):
% - - - - - - - - - - - - - - - - - - - - - - - - - - - - - - - - - - - %
The analog of route (1) is to start with the geometrical dual
of the $(p,q)$ five-brane \reIIB. This is given in terms of
M theory on $X_{p,q}=(\MR{4}\times\MS{1})/\BZ_q$ where the $\BZ_q$ acts
on $\MR{4}$ to give an $A_{q+1}$ singularity and on $\MS{1}$ by
a shift. The decoupling limit requires taking the size of $\MS{1}$
to zero.
The matrix model for this M theory background follows immediately from
studying zero-brane dyanmics on $X_{p,q}$ \reIIB. It is a 1+1D gauge theory
on $\MHS{1}$ with gauge group,
$$
U(N)_1\times\cdots\times U(N)_q,
$$
and cyclic hypermultiplets. The integer $p$ enters as a twisted boundary
condition along $\MHS{1}$ for the global $\BZ_q$ symmetry which
rotates the chain.
To compactify on $\MT{d}$, we take a $(d+2)$-dimensional
gauge theory with the same field content compactified on
$\MHT{d}\times\MHS{1}$,
with boundary conditions twisted by the global $\BZ_q$
along $\MHS{1}$. For sufficiently large $d$, the description in terms of a
field
theory will break down and we should look for a decoupled theory to define the
matrix model. In this case, the desired string theories are easily related to
\rkin. However, in other cases, for example those related to non-simply-laced
gauge groups, there are theories with novel properties. We hope to report on
these models elsewhere.

% - - - - - - - - - - - - - - - - - - - - - - - - - - - - - - - - - - - %
% Route (2):
% - - - - - - - - - - - - - - - - - - - - - - - - - - - - - - - - - - - %
In route (2), we start with M theory compactified on $\MT{2}$
and add a longitudinal five-brane wrapping a $(p,q)$ 1-cycle.
The matrix model for this case is given by 2+1D SYM compactified
on $\MHT{2}$ with a line of impurities along a $(p,q)$ 1-cycle.
In a fundamental cell, this looks like ${{pq}\over {r}}$ parallel
lines of impurities. After compactification on a further $\MHT{3}$,
we obtain what at first sight looks like
the $(0,q)$ theory compactified on $\MHT{4}\times\MS{1}$ with
${{pq}\over {r}}$ parallel lines of impurities.
As in the previous discussion,
we are taking the limit where $\MS{1}$ is much bigger than $\MHT{4}$.
The phrase `lines of impurities' has to be translated into the
phrase `the sector with $q$ units of string winding along
$\MS{1}$ and $p$ units of winding along one of the other
directions of $\MHT{4}$.'
In terms of the previously defined 1+1D theory $\X_{q,N}$,
this is interpreted as follows. $\X_{q,N}$ has 8 conserved
quantum numbers associated to momentum and winding along $\MT{4}$.
Each of them is related to a $U(1)$ symmetry corresponding to
a shift in an approprite direction in $\MT{4}$, or after T-duality $\MHT{4}$.
 We pick a $\BZ_q$ subgroup of
one of these $U(1)$ symmetries
and take $X_{q,N}$ on $\MS{1}$ with a twisted boundary condition
with respect to this subgroup. Heuristically, the effect of this
twist is to close the $q$ strings along $\MS{1}$ up to a shift,
which exactly agrees with the impurity picture.

% - - - - - - - - - - - - - - - - - - - - - - - - - - - - - - - - - - - %
% Route (3):
% - - - - - - - - - - - - - - - - - - - - - - - - - - - - - - - - - - - %
In route (3) we start with the $(p,q)$ theory compactified
on a small $\MS{1}$ with $N$ units of KK momentum. We then perform
T-duality along this $\MS{1}$ and another direction of the $\MT{4}$.
We end up with the $(0,q)$ five-brane theory compactified on another
$\MT{4}$ with $q$ units of winding as before. The new ingredient
is that the $p$ charge becomes magnetic flux along the $\MHS{1}$
and another direction of $\MT{4}$.
We see again that the $(p,q)$ theory is related to twisting
by an element of a global $\BZ_q$ symmetry of $\X_{N,q}$.
This time the $\BZ_q$ symmetry in question is a gauge transformation
by a topologically nontrivial gauge transformation of $SU(q)$.
This gauge transformation corresponds to a nontrivial generator
of $\pi_1(SU(q))$ \rtHooft.

% =====================================================================
%
% Matrix Models for Yang-Mills Theories
% =====================================================================
%

\newsec{Matrix Models for Yang-Mills Theories}

% --------------------------------------------------------------------- %
% Matrix theory on an ALE space
% --------------------------------------------------------------------- %

We are now ready to take the IR limit in various dimensions to obtain matrix
models
for SYM theories. We start with M theory on $T^d \times {\rm A_{k-1}}$ where
the radii of the torus are  $ \{ R_1, \ldots, R_d \}$. The effective SYM
coupling
constant is given in \yangmillscoupling. The matrix model, given in section
{\it 2.2}, has
eight supersymmetries and lives on the dual torus, $\MHT{d}$, with sides
$$
\Sigma_j = {1\over {M_{pl}^3 \RV R_j}}.
$$
The matrix model coupling is,
\eqn\matrixcoupling{ (g_{d+1})^2 = {1\over {M_{pl}^{3(d-2)} \RV^{d-3} R_1
R_2\dots R_d}},}
and for $d<4$, the matrix model is a gauge theory. The scale of the matrix
theory is
set by the effective coupling constant and for $d=3$, the theory is conformal.

We will show below that the limit in which the matrix theory describes
Yang-Mills
corresponds to reducing the degrees of freedom of the full matrix model to
quantum
mechanics on the Coulomb branch. We restrict ourselves to
energies below the compactification scale,
\eqn\kkscale{  E\ll R_1^{-1},\dots,R_d^{-1},}
with the spacetime coupling $\hat{g}_{d}$ held fixed.
A word on notation: we will occasionally take a limit like,
$$
R_j \rightarrow 0,
$$
which is shorthand for ``we keep our energy scale much smaller than
$R_j^{-1}$.''
There are now two issues that we need to address: the first is whether the
energy scale set by the sides $\S_j$ of $\MT{d}$ is small compared to the scale
set by
the coupling. If so, we can first solve for a desciption of the low-energy
physics using
the dynamics of $d+1$-dimensional
SYM. The
second issue is whether the energy scale we wish to observe is smaller than
$\S_j^{-1}$,
in which case, we can further reduce to quantum mechanics on the moduli space.
We will consider this reduction case by case.

% --------------------------------------------------------------------- %
% 4+1D SYM
% --------------------------------------------------------------------- %

\subsec{Five-dimensional SYM}

To obtain 4+1D SYM, we consider M theory on $\MT{2} \times A_{k-1}$. As we
discussed
in the previous section, this can be viewed as compactifying the $(1,1)$ string
theory
on a circle. The classical moduli space can be parametrized by Wilson lines
along
the compact circles,
\eqn\classical{\CM =  \left[ (\IR^3 \times T^2)^N/{\bf S}_N \right]^k, }
while the metric generally receives quantum corrections. The dynamics is
governed by the
parameters,
$$ \eqalign{(g_3)^2 \S_1 & = {1\over M_{pl}^3 R_1^2 R_2} = {(\hat{g}_5)^2 \over
R_1} \gg 1\cr
(g_3)^2 \S_2 & = {1\over M_{pl}^3 R_1 R_2^2} = {(\hat{g}_5)^2 \over R_2} \gg 1.
}$$
We can therefore reduce to the $2+1$-dimensional moduli space parametrized by
the Wilson
lines and the scalars in the vector multiplets. Restricting to energies obeying
\kkscale\
and taking $M_{pl} \r \infty$ implies that our energy scale satisfies,
$$ E \ll {1\over \S_1}, {1\over \S_2}.$$
We can therefore neglect the dependence on the coordinates $x_1, x_2$ and
reduce to quantum
mechanics on the moduli space.

The coupling in the quantum mechanics,
\eqn\quantumcoupling{ g^2_{QM} = {\RV^3 M_{pl}^6},}
goes to infinity. There are two energy scales associated to wavefunctions on
\classical\ with
non-constant dependence on the two compact directions. The energy scale for the
direction
coming from a Wilson line along $x_1$ is,
$$ E \sim {\RV \over R_1^2} \r \infty. $$
We can then take our wavefunctions to be locally independent of the coordinate
coming from $A_1$.
Likewise, the other compact direction has an energy scale,
$$ E \sim {\RV \over R_2^2}, $$
which becomes large as we take the five-dimensional limit, $R_2 \r 0$. We can
then
restrict to ground state wavefunctions on the compact directions, with a
non-trivial dependence on the three non-compact directions. This limit
describes the infra-red dynamics of five-dimensional
Yang-Mills.

There is a second way to obtain this result which is essentially T-dual to the
previous
description and closely related to the field theory limit of our discussion in
section
{\it 2.2}. We can define 4+1D SYM as the $(2,0)$ theory compactified on
$\MS{1}$
 \rfourtorus. The size of the circle is $ \hat{g}_5^2$. In this case, we
parametrize the
Coulomb branch in terms of the scalars dual to the photons and the three
scalars in a
vector multiplet. Including quantum corrections, the Coulomb branch is a
hyper\kh
metric on $ \left[\IR^3 \times S^1)^N / {\bf S}_N \right]^k $ where the circles
have
size $ \hat{g}_5^2 $. In the limit $ \hat{g}_5^2 \r \infty$, the manifold
reduces to the
moduli space of $N$ instantons in $SU(k)$ gauge theory, and we recover our
description
of the $(2,0)$ field theory.

We then have a $2+1$-dimensional sigma model from $\MHT{2}$ to the Coulomb
branch.
We would like to see the full compactified $(2,0)$ theory so we need to
consider
energies associated to the dual photons,
$$ E \sim \RV M_{pl}^6 (R_1 R_2)^2. $$
The scale of fluctuations for the non-compact scalars, $1/\S_i$, is much larger
than
the energies under consideration. The same is true for the scale of
fluctuations for
the dual photons for the same reason that we could neglect Wilson lines in the
preceeding discussion. We can therefore restrict to quantum mechanics on the
hyper\kh moduli space. In the IR limit, we can restrict to ground state
wavefunctions
along the compact directions, which means that the wavefunctions are locally
constant
in the compact variables.

Since the spectrum includes massless vector bosons, we should have $L^2$
zero-energy ground states in the quantum mechanical sigma model for every
 $N$. With this amount of supersymmetry, the ground states should correspond to
 elements
of de Rham cohomology. For the case of $N=1$, it should be possible to check
the
 existence of these forms using the metric presented in \ryi. For example, for
$k=2$, the metric has an $A_1$ singularity and the desired $L^2$ form
comes from this singularity. Similar forms should exist for every $N$ and $k$.

% --------------------------------------------------------------------- %
% 3+1D SYM
% --------------------------------------------------------------------- %

\subsec{Four-dimensional SYM}

The physics is considerably different for $d=3$. The $U(1)$ part of the $U(N)$
factors
 in \firstgauge\ are not asymptotically free. The effective gauge group in the
infra-red is then,
\eqn\secondgauge{ SU(N) \times \ldots \times SU(N).}
Note that for $d<3$, we require the gauge group given in \firstgauge\ or for
example,
T-duality of the string theory would fail.  The matrix theory in
four-dimensions
is then a finite theory with coupling,
\eqn\threecoupling{ g_4^2 = {1 \over M_{pl}^2 R_1 R_2 R_3}.}
It is a requirement for this proposal to be sensible that the coupling be a
true modulus
 of the theory. If the matrix theory were not a finite gauge theory, it almost
certainly could not be describing an S-dual theory. This further supports the
choice \secondgauge. For this reason, the matrix models for the $D_k$ and $E_6,
E_7, E_8$ cases must also be finite theories, and indeed they are finite.
Again, the
$U(1)$ factors and associated charged matter are frozen out.

The classical moduli space is now,
\eqn\newmoduli{ \CM  =   \left[ (\IR^2 \times T^3)^{N-1}/{\bf S}_{N-1}
 \right]^k. }
The spacetime coupling and the matrix model coupling are identical, and we will
hold \threecoupling\ fixed. Because the coupling is dimensionless, the dynamics
is independent of the size of $\MT{3}$. At sufficiently low energies,
we can always reduce to quantum mechanics
on the moduli space. By low energies, we mean energies far below
$\Sigma_j^{-1}$.
To see this let us again estimate the relevant energy scales. Let us normalize
the Lagrangian for a scalar field $\phi$ on the moduli space to be,
$$
L = \int d^3 x dt |\px{\u}\phi|^2.
$$
Truncating to quantum mechanics keeps only the zero mode for $\phi$ giving,
$$
L = \Sigma_1 \Sigma_2 \Sigma_3 \int dt |\dot{\phi}|^2.
$$
For energies $E \ll \Sigma_j^{-1}$, we can estimate the spread of the
wavefunction $\psi (\phi)$ to be,
$$
\Delta\phi \sim E^{-1/2} (\Sigma_1 \Sigma_2 \Sigma_3)^{-1/2}.
$$
On dimensional grounds, higher derivative terms in the low-energy
expansion will have more powers of $\phi^{-1}\px{\u}$. We can estimate the size
of these corrections:
$$
\phi^{-1}\px{j} \sim E^{1/2} \left(
{{\Sigma_1 \Sigma_2 \Sigma_3}\over {\Sigma_j^2}}\right)^{1/2}
\sim (E\Sigma_j)^{1/2} \ll 1.
$$
We can then restrict to ground state wavefunctions on the torus, so
finally  restrict to quantum mechanics on the Coulomb branch of the quiver
gauge theory.

% - - - - - - - - - - - - - - - - - - - - - - - - - - - - - - - - - - - %
% Correspondences
% - - - - - - - - - - - - - - - - - - - - - - - - - - - - - - - - - - - %

It is interesting to derive this result from a limit of
the $(2,0)$ theory compactified on $\MT{2}\times\MWT{2}$.
We take the limit where the complex structure of  $\MWT{2}$ is fixed to be
$\tau$, and where the volume goes to zero:
\eqn\mwtz{
\vol(\MWT{2})\longrightarrow 0.
}
The other $\MT{2}$ is taken to be the transverse space for $\SUSY{4}$
SYM. The matrix model for this
theory is described by quantum mechanics on the moduli space of
$U(N)$ instantons on the dual $\MHT{2}\times\MHWT{2}$
at instanton number $k$. This moduli space has $4 N k$ real dimensions.
In the limit \mwtz\ this moduli space
has large and small directions. Quantization of the motion along
the small directions gives a very high energy scale which is related
to the compactification scale of $\MWT{2}$. To get 3+1D SYM, we need
to be far below this scale. This means that we are only interested
in wavefunctions which do not vary locally along the small directions
of the moduli space. There are
$2 N k + 2$ large directions and $2 N k -2$ small ones.
To see this we note that in the limit where
$\MHT{2}$ is much smaller than $\MHWT{2}$,
instantons on $\MHT{2}\times\MHWT{2}$ are characterized by specifying
the $U(N)$ holonomies on the small $\MHT{2}$. These are given
by $N$ points on the dual of $\MHT{2}$ which brings us back to
$\MT{2}$. This divisor of $N$ points on $\MT{2}$
has to vary holomorphically over $\MHWT{2}$.
It traces a Riemann surface $\Sigma\subset \MT{2}\times\MHWT{2}$.
It is an $N$-fold cover of $\MHWT{2}$ and it is also easy
to see that $\Sigma$ is a $k$-fold cover of $\MT{2}$.
It is not hard to compute the genus of such a curve and we find
that $g = k N +1$. The moduli space of such curves has $2 k N + 2$
(real) dimensions which are the large directions.
It is also possible to identify the small dimensions.
It actually true that the $(4 N k)$-dimensional
moduli space ${\cal M}_{N,k}$
of instantons on  $\MHT{2}\times\MHWT{2}$ for any size $\MHT{2}$
is fibered over the $(2 N k +2)$-dimensional moduli space of curves
inside $\MT{2}\times\MHWT{2}$. This is a special case of the
{\it spectral curve} construction (see \rFMW\ for details) and
the remaining data that needs to be given is a line-bundle over
the curve $\Sigma$, which is the spectral bundle.  We will explain in a
later section why the dimension of this moduli space differs from the
dimension of the Coulomb branch.

% --------------------------------------------------------------------- %
% 2+1D and 1+1 SYM
% --------------------------------------------------------------------- %
\subsec{Low-dimensional SYM}

The matrix models for $d>3$ are ill-defined since the field theories are
not asymptotically
free.  For d=4, the
matrix model coupling is,
\eqn\fourcoupling{ g_5^2 = { 1 \over M_{pl}^4 \RV R_1 R_2 R_3 R_4},}
and determines a length scale in the problem. The moduli space of
$4+1$-dimensional
Yang-Mills has a metric which is linear in the moduli \rSeiFDS. We need to
check
whether at energies obeying \kkscale, the dynamics is determined by the UV
or IR behavior of the
theory. We compare,
$$ {g_5^2 \over \S_i} = (\hat{g}_3)^2 R_i \r 0,$$
which implies that the effective coupling is weak at the scale set by
$\MHT{4}$. By
similar reasoning to the previous cases, we can then reduce to quantum
mechanics
on the moduli space of the $4+1$-dimensional theory. After reducing to quantum
mechanics, it is natural to rescale the moduli absorbing the volume of
$\MHT{4}$
and the coupling constant. In terms of the rescaled variables $\tilde{\phi}$,
the
kinetic term is of the form:
\eqn\metric{ (1 + \RV (\hat{g}_3)^4 \tilde{\phi}) ({d \tilde{\phi} \over
dt})^2.}
This quantum mechanics describes the physics below energies of order $1/R_i$
but
includes energies around $\hat{g}_3^2$.

At first sight, in the conformal limit, it seems that the second term in
\metric\
dominates. However, on closer inspection, we see that the moduli space has
singularities at finite distance. For example, in the case of $SU(2)_1\times
SU(2)_2$
with hypermultiplets, the two effective gauge couplings $h_1$ and $h_2$ depend
on
the two scalars, $\phi_1$ and $\phi_2$, in the vector multiplets in the
following way:
$$ {1\over h_1^2} = {1\over g_5^2} + |\phi_1| - |\phi_2|,\qquad
  {1\over h_2^2} = {1\over g_5^2} - |\phi_1| + |\phi_2|.$$
It seems that for the metric to remain positive definite, the combination
$ |\phi_1| -|\phi_2|$ is bounded by $1/g_5^2$. If true, then the low-energy
dynamics
is described by the combination $ |\phi_1| +|\phi_2|$ with a flat metric. This
is
quite puzzling and we will return to this point in section four after we
discuss the
DLCQ Wilson line.

For the case of $d=5$, we need to understand the low-energy behavior of NS
five-branes
at the $A_{k-1}$ singularity \rjulie\ which is quite difficult. Instead, we
will
discuss the matrix model for 1+1D SYM from a different route in the
following section.

% =====================================================================
%
% Section (4) : Features of the Matrix Description
% =====================================================================
%

\newsec{Exploring Features of the Matrix Description}

Now that we have arrived at the matrix model for 3+1D $\SUSY{4}$ SYM
from various viewpoints, we are ready to explore how the known
properties of SYM are visible in this model. The main advantage of
this DLCQ model is that S-duality is manifest. It is therefore
interesting to see how perturbative features of SYM emerge.

% --------------------------------------------------------------------- %
% Summary of the previous sections
% --------------------------------------------------------------------- %
\subsec{Summary of the model}

We have argued that the $p_\Vert = N$ DLCQ sector
of 3+1D $SU(k)$ SYM with both transverse directions
compactified on $\MT{2}$ is given by quantum mechanics on
a certain curved manifold ${\wM_{N,k}}$ of (real) dimension
$ 2 N k$. The space
$\wM_{N,k}$ is described as follows:
start with the product space $\MT{2} \times \MWT{2}$ where $\MWT{2}$ has
complex structure $\tau$. In the Coulomb branch approach, we only saw the limit
$\MT{2} \r \IR^2$. Let $\wM'_{N,k}$ be the moduli space of holomorphic
Riemann surfaces $\S$ in $\MT{2}\times\MWT{2}$ which intersects $k$ times any
$\{p\} \times \MWT{2}$ with $p$ a generic point in the transverse space, and
$N$
times any
$\MT{2}\times \{p'\}$ again for generic $ \{p'\}$. This space $\wM'_{N,k}$ has
dimension
$2Nk +2$ and admits a torus action ($U(1)\times U(1)$) which
translates the curve $\S$ along $\MWT{2}$. To obtain ${\wM_{N,k}}$, we quotient
$\wM'_{N,k}$ by this torus action. Let
$$
q^1\dots q^{g-1},\qquad g \equiv kN + 1
$$
be local complex coordinates on the moduli space $\wM_{N,k}$
of such surfaces.
Locally, near a point $z\in \MT{2}$, the curve can be described
by $k$ functions from the neighborhood of $z$ into $\MWT{2}$:
$$
w_r(z,q):\MT{2}\longrightarrow\MWT{2}\qquad r=1\dots k
$$
and $q$ is a shorthand for $q^1\dots q^{g-1}$.
These functions define a Riemann surface $\Sigma\subset \MT{2}\times\MWT{2}$
of genus $g = k N +1$.
The metric for the 0+1D Lagrangian $ \left( \int {1\over 2} g_{i\bar
{\jmath}}\dot{q}^i
\dot{\widebar{q}}{}^{\bar{\jmath}} + \ldots \right)$ can be written as
\eqn\kinagain{
g_{i\bar{\jmath}}
=  \RV^{-1} \sum_{r=1}^k
\int\ d^2 z \, \left(\pypx{w_r}{q^i}\cdot \pypx{\widebar{w}_r}{\widebar{q}^{
\bar{\jmath}}} \right).
}
We have normalized the area of $\MWT{2}$ to be one. The local 1-forms,
$$
\pypx{w_r}{q^i} dz,\qquad i=1\dots g-1,
$$
can be patched to a global holomorphic 1-form $\mu_i$ on $\Sigma$. So
we can write,
$$
g_{i\bar{\jmath}} =  \RV^{-1} \int \mu_i\wdg\widebar{\mu}_{\bar{\jmath}}.
$$
Under a complex supersymmetric variation $ \delta$, we see that
$$
\delta q^i = \epsilon_a \psi^{i a}, $$
The holomorphic spinors $\psi^{i a}$ have spin-$\half$ under the
local rotation group of the transverse space $\MT{2}$ and the index
$a=1\dots 4$ is a spinor of the $SO(6)$ R-symmetry.
As usual, there are quadratic and quartic fermion terms which provide
the usual supersymmetric completion of the sigma model.

% --------------------------------------------------------------------- %
% DLCQ Wilson line
% --------------------------------------------------------------------- %
\subsec{DLCQ Wilson line}

When we take the limit $\MT{2} \r \MR{2}$, the vacua of the system
 should be characterized
by the value of the $U(k)$ Wilson line
along the light-like direction.
In order to see this variable in the matrix description, we note that
as $\MT{2} \r \MR{2}$ the Riemann surfaces
$\Sigma\subset \MR{2}\times\MWT{2}$ are no longer compact.
We now have to specify the boundary condition at the boundary of $\MR{2}$.
These boundary conditions are $k$ limiting points on $\MWT{2}$.
The projection of these points on one axis of $\MWT{2}$
is in one-to-one correspondence with the conjugacy class of the $U(k)$
Wilson line which is specified by a point in $(\MS{1})^k/S_k$.
The projection on the other axis corresponds to `magnetic'
Wilson lines or, in other words, the VEV of the dual photon since the
system is effectively 2+1D.

In this limit, we then freeze $2k-2$ (real) moduli. The original moduli space
$\wM_{N,k}$ had dimension $2Nk$. The final moduli space then has $ 2k(N-1)+2$
parameters while on the other hand, the Coulomb branch picture led to a moduli
space with dimension $2k(N-1)$. The discrepancy is accounted for by noting that
the two parameters correspond to translation of the curve along $\MR{2}$. In
the
Coulomb picture, this center of mass motion corresponds to the diagonal $U(1)$
factor from \firstgauge\ which had no charged matter, and therefore remains a
modulus in 3+1D.

% --------------------------------------------------------------------- %
% Electric and magnetic fluxes
% --------------------------------------------------------------------- %
\subsec{Electric and magnetic fluxes}

The curves $\S$ becomes singular on a subspace $H_{N,k}$. In the Coulomb branch
picture,
the singular locus corresponds to the points where the Higgs and Coulomb
branches meet.
What boundary conditions should we impose on the wavefunctions at the singular
locus
$H_{N,k}$? We will provide a partial answer to this question.

We want to argue that $\pi_1 (\wM_{k,N} - H_{N,k})$ is non-trivial, and
contains
$ \IZ_k \times \IZ_k$. To see this, note that $\wM'_{k,N}$ has paths connecting
two
points which are identified under the torus action which gives $\wM_{k,N}$.
Intuitively,
this path is given by smoothly varying the moduli and couplings of the $U(N)_i$
factor
into those of the $U(N)_{i+1}$ factor cyclically around the chain \firstgauge.
This means
that we can consider sectors where the wavefunctions pick up a phase in
$ \IZ_k \times \IZ_k$ when taken around this loop in moduli space. To what can
these boundary
conditions correspond?

3+1D $SU(k)$ SYM compactified on $\MT{3}$ has
different sectors specified by the discrete $\BZ_k$ electric and
magnetic fluxes along the sides of $\MT{3}$ \rtHooft.
If the gauge group were $U(k)$ instead of $SU(k)$, there would also be fluxes
for the center $U(1)$ subgroup. The flux for the $SU(k)\subset U(k)$
is correlated with the $U(1)$ flux and the sector with
$p\in\BZ_k$ units of $SU(k)$ flux, either electric or magnetic, has to have
a $U(1)$ flux in $\BZ + {p\over k}$ in the same direction.
The $U(k)$ fluxes are characterized by the $U(1)$
fluxes in $\BZ$ except that the quantization is in units of ${1\over k}$.
In the DLCQ, we put $U(k)$ SYM on $\MT{2}\times\MS{1}$ where
$\MS{1}$ is light-like.

Let $\MT{2}$ be with sides of sizes $R_1,R_2$ and $\MS{1}$ is, as usual,
of size $\RV$.
In principle,
we can have electric and magnetic fluxes either in the $\MS{1}$
direction, or in the $\MT{2}$ direction, or both.
In the DLCQ, we can only observe the fluxes in the direction of $\MS{1}$.
Fluxes in the transverse directions have an energy
proportional to the inverse of the invariant length of $\MS{1}$,
which becomes infinite as $\MS{1}$ becomes light-like.
Another way to see this is to note that electric fluxes in a certain
direction are the canonical dual variables to Wilson lines in this
direction. In the DLCQ, this Wilson line is like a zero mode of a field
(i.e. with $p_\Vert = 0$) and we cannot quantize the zero modes in the
DLCQ. We claim that the twisted sectors where the wavefunction picks up a
 phase $e^{2\pi i (n_1/k)}$
for one generator of $\IZ_k$ and $e^{2\pi i (n_2/k)}$ for the second generator
 correspond to
electric and magnetic fluxes along the DLCQ direction.

Let us consider what changes if the gauge group were $U(k)$ rather than
$SU(k)$. We would
not be able to see the propagating modes for the $U(1)$ factor in this
approach, but since
the propagating $U(1)$ modes decouple from the $SU(k)$ theory, we do not lose
much. However,
the change to $U(k)$ does change the global structure by shifting the energy of
the twisted
sectors. The sectors with different fluxes are no longer degenerate, but have
energies
shifted by an amount proportional to,
\eqn\energy{
{1\over k}\tau_2 {{|n_1 + n_2 \tau|^2}\over {R_1 R_2}}.
}
What changes in the matrix model construction? From route three described in
section two,
we can see that this requires replacing $\wM_{N,k}$
by $\wM'_{N,k}$. The change amounts to introducing two extra collective
coordinates which
correspond to translating the curves $\S$ along $\MWT{2}$. Quantizing these
extra collective coordinates gives the energy levels in \energy.

% --------------------------------------------------------------------- %
% W-bosons in the perturbative limit
% --------------------------------------------------------------------- %
\subsec{W-bosons in the perturbative limit}

How do we see the W-bosons in the matrix model?
Since we are at the origin of the moduli space, the theory
is conformal and strictly speaking there are no asymptotic states.
Therefore $W^0$ and $W^\pm$ do not make sense as particles.

Since we have a generic Wilson line in the DLCQ,
the charged $W$-particles become effectively massive and
we can try to look for them.
For simplicity let us work with $SU(2)$.
A DLCQ Wilson line can be specified by $0 < \a < 1$.
In the presence of the Wilson line, charged particles do not
have integral $\pV$ anymore (see also \rHelPol).
The $W^+$ particles can have
$$
\pV = \a, 1 + \a, 2+\a, \ldots
$$
while the $W^-$ particles can have
$$
\pV = 1-\a, 2-\a, 3-\a, \ldots.
$$
In the presence of the Wilson line, the sector with $\pV = 2$,
for example, can have 4 particles: two $W^+$ particles with $\pV = \a$
and two $W^-$ particles with $\pV = 1-\a$.

% - - - - - - - - - - - - - - - - - - - - - - - - - - - - - - - - - - - %
% The perturbative limit.
% - - - - - - - - - - - - - - - - - - - - - - - - - - - - - - - - - - - %

It is easier to search for the $W$ particles in the perturbative limit.
In the perturbative limit, the auxiliary torus $\MWT{2}$ becomes
very elongated. To see what the curves $\Sigma$ look like, it is most
useful to recall that these are the Seiberg-Witten curves of
the $\SUSY{2}$ $SU(N)_1\times SU(N)_2$ gauge theory with
matter in the $(N,\widebar{N})$ and $(\widebar{N},N)$ and coupling
constants,
\eqn\aparameter{
\tau_1 = \tau \a,\qquad \tau_2 = \tau (1-\a),
}
where for simplicity we have set the $\theta$ angles to zero. For the same
reason, let us set $N=2$.
In the perturbative limit, the moduli space is parameterized
by the VEVs of the two vector multiplets
$\pm z_1$ and $\pm z_2$, respectively.
The leading term in the metric over the moduli space is
the classical term which implies the following kinetic energy
for the matrix model:
$$
\a |\dot{z}_1|^2 + (1-\a) |\dot{z}_2|^2.
$$
We interpret this as two $W^+$-bosons with $\pV = \a$
at transverse positions $z_1$ and $-z_1$ and two $W^-$-bosons
with $\pV = 1-\a$ at transverse positions $\pm z_2$.

% --------------------------------------------------------------------- %
% Coulomb interaction of W-bosons
% --------------------------------------------------------------------- %

\subsec{Coulomb interaction of W-bosons}

We would also like to reproduce the interaction between W-bosons.
It is well-known that the force between two $W^+$-bosons
is velocity dependent. The static force generated by the exchange of
photons and Higgs particles cancel.
However, the static force between a $W^+$ and a $W^-$-boson
is not cancelled and it is interesting to see how this force is reproduced in
the
matrix model.

When a positively charged particle
 and a negatively charged particle are both parallel and moving toward each
other
at the speed of light, the electric and magnetic interactions
exactly cancel. Thus, in the DLCQ the electro-magnetic
potential energy between two particles with charges
$Q_1$ and $Q_2$, at transverse positions
$z_1$ and $z_2$, is given by:
$$
V\sim Q_1 Q_2 |\dot{z}_1 - \dot{z}_2|^2 \log |z_1 - z_2|.
$$
Note that $\log |z_1 - z_2|$ is the solution to Poisson's equation
in the transverse two dimensions and that the expression
$|\dot{z}_1 - \dot{z}_2|^2$ is Galilean invariant.
After adding the Higgs exchange contribution to this expression,
we find that the potential energy for a pair of $W^+$-bosons, or
a $W^+$ and a $W^-$ are both proportional to:
$$
|\dot{z}_1 - \dot{z}_2|^2 \log |z_1 - z_2|.
$$
%The energy for two $W^{+}$ bosons is larger by a factor of $(-2)$
%than the energy for a $W^{+}$ and a $W^-$.

% - - - - - - - - - - - - - - - - - - - - - - - - - - - - - - - - - - - %
% Matrix interpretation
% - - - - - - - - - - - - - - - - - - - - - - - - - - - - - - - - - - - %
In the matrix model, these expressions should be obtained as
corrections to the metric on the moduli space.
Viewed as the moduli space of $SU(N)_1\times SU(N)_2$ gauge theory where $N=2$,
the first-order correction is a 1-loop effect which indeed behaves
like,
\eqn\onelp{
|\dot{z}_1 - \dot{z}_2|^2 \log |z_1 - z_2|,
}
where $z_1$ and $z_2$ are VEVs for the scalar partners of the
gauge fields; we are in the perturbative limit here.
The difference between the energy of two like particles
and the energy of opposite particles is accounted for by noting
that for two $W^+$ bosons, both $z_1$ and $z_2$ correspond to
scalar VEVs of the same $SU(N)_1$ factor and the term \onelp\
comes from a loop containing the vector multiplet and bifundamentals.
For a $W^+$ and a $W^-$, $z_1$ and $z_2$ belong to different $SU(N)$
factors and \onelp\ is the effect of a hypermultiplet.

% --------------------------------------------------------------------- %
% Enhanced R-symmetries
% --------------------------------------------------------------------- %
\subsec{Enhanced R-symmetries}

So far we have dealt with perturbative effects.
We now turn to a non-perturbative effect which appears at strong coupling.
This is the enhanced $Spin(8)_R$ symmetry in the IR limit of 2+1D SYM
with $\SUSY{8}$.

Let us recall some facts about R-symmetry in theories with
16 supersymmetries (see \rreview\ for more details).
R-symmetry is a global symmetry which acts on the
supersymmetry generators.
In theories with $\SUSY{4}$ in 3+1D the maximal R-symmetry possible
is $Spin(6) \equiv SU(4)$ since we have 4 complex generators.
For 3+1D gauge theories this is indeed a symmetry and the 6 scalars
transform in the $\rep{6}$ of $Spin(6)$.
On the other hand, if we take the 5+1D $(2,0)$ theory and compactify
on $\MT{2}$, we obtain a 3+1D theory with 16 supersymmetries in the
uncompactified dimensions. This theory only has $Spin(5)$ R-symmetry.
Its moduli space is not $\MR{6}$ but $\MR{5}\times\MS{1}$ and there
is no $Spin(6)$ symmetry. In the low-energy limit, however, the full
$Spin(6)$ is restored. Similarly, 2+1D SYM with $\SUSY{8}$ has
a manifest $Spin(7)$ R-symmetry. In the IR limit, this theory flows
to a nontrivial fixed point with an enhanced $Spin(8)$ R-symmetry.
We could have also started with either the $(1,1)$ or $(2,0)$ 5+1D string
theories
compactified on $\MT{3}$. This theory generically has only a $Spin(4)$
R-symmetry but in the IR limit has an enhanced  $Spin(8)$ R-symmetry.
Similarly, the 5+1D theory on $\MT{2}$ has generically an $Spin(4)$
symmetry which is enhanced to $Spin(6)$ in the IR limit.
The $Spin(6)$ symmetry of 3+1D SYM is, of course, a perturbative feature
but the enhanced $Spin(8)$ of 2+1D SYM at low-energies is
non-perturbative.

In the matrix model, we can see some of these phenomena to a greater or
lesser extent.
We have argued that the model for the non-critical string theory
compactified on $\MT{2}$ is generically some 1+1D CFT at low-energies.
It is obtained by compactifying a 5+1D theory on $\MT{4}$.
Generically, the $Spin(4)$ rotation symmetry of the 5+1D theory
in the directions of the $\MT{4}$ is broken.
However, we have seen in section (2) that under certain
conditions the matrix model for the
low-energy limit of the 3+1D or 2+1D
theory is given by the low-energy limit of a 3+1D theory compactified
on $\MT{3}$, or a 4+1D theory compactified on $\MT{4}$, respectively.
In the case of 2+1D theories with 16 supersymmetries,
we obtained a 4+1D matrix model compactified
on $\MT{4}$. In the low-energy
limit the size of the $\MT{4}$ was very large compared to the
scale of the theory and we could
reduce to quantum mechanics on the moduli space.
In the IR limit, the $Spin(4)$ rotation symmetry in the directions of
$\MT{4}$ was restored even though the full spectrum
(which was sensitive to the finite size of the $\MT{4}$) was not
$Spin(4)$ symmetric. This $Spin(4)$ symmetry acts only on the fermionic
variables of the quantum mechanics.
The fermions also have quantum numbers under the $SU(2)$ R-symmetry
of the 4+1D theory. This gives an obvious
$Spin(4)\times Spin(3)\subset Spin(7)$ symmetry.
Inspection of the Lagrangian reveals that the symmetry is the full
$Spin(7)$. This is because every fermionic variable is in the
$$
(\rep{2},\rep{1})\times\rep{2} + (\rep{1},\rep{2})\times\rep{2}
$$
of $Spin(4)\times Spin(3)$
and thus comes with a multiplicity of 8.
Let $\lam^{\a i}$ be a fermionic
variable with $\a=1, 2$ and $i=1, 2$ and let
$\lam^{\dot{\a} i}$ be the other variables.
The kinetic energy terms are proportional to
$$
\epsilon_{ij}\epsilon_{\a\b}\lam^{\a i}{\partial_t}{\lam}^{\b j}
+
\epsilon_{ij}\epsilon_{\dot{\a}\dot{\b}}\lam^{\dot{\a}
i}{\partial_t}{\lam}^{\dot{\b} j}
$$
which is $Spin(8)$ symmetric.
The quartic terms on the other hand look like
$$
 D^m D^m,\qquad m=1\dots 3,
$$
where
$$
D^m =
\epsilon_{\a\b}\sigma^m_{ij}\lam^{\a i}\lam^{\b j}
-\epsilon_{\dot{\a}\dot{\b}}\sigma^m_{ij}\lam^{\dot{\a} i}\lam^{\dot{\b} j}.
$$
A tedious calculation shows that $\lam^{\a i}$ and $\lam^{\dot{\a} i}$
can be combined into a spinor $\lam$
of $Spin(7)$ which enlarges $Spin(4)\times Spin(3)$ and
$$
 D^m D^m = (\lam\Gamma^A\lam)(\lam\Gamma_A\lam),\qquad
A=1\dots 7,
$$
where $\Gamma^A$ are Dirac matrices of $Spin(7)$.
Thus the R-symmetry of the quantum mechanics is indeed $Spin(7)$.
It cannot be $Spin(8)$ since $Spin(8)$ does not
have an anti-symmetric quartic invariant as can be seen by triality.

How does $Spin(8)$ appear in the IR limit of the 2+1D SYM? We found in
section {\it 3.3} that the IR limit seemed to be described by a flat metric.
In this case, the $Spin(8)$ violating curvature term vanishes and we trivially
recover a $Spin(8)$, but not the $Spin(8)$ of the spacetime SYM theory. A
probable
resolution to this puzzle is that we are describing a model with a DLCQ Wilson
line.
The Wilson line distinguishes the dual photon from the other 7 scalars, and so
breaks
$Spin(8)$. With the Wilson line, the theory will now flow to a free theory in
the IR rather
than the interacting fixed point. To recover a model with $Spin(8)$, we need to
describe the
 case without
a DLCQ Wilson line. A possible way to turn off the Wilson line is to
tune $\a$ in \aparameter\ to zero, which takes one of the bare couplings to
infinity.
We will not explore this possibility further here.

Similarly, the 3+1D theory had a matrix model which was a 3+1D
theory compactified on $\MT{3}$. In the IR limit, the rotation
group $Spin(3) \equiv SU(2)$ of $\MT{3}$ was restored.
The fermionic variables in the quantum mechanics now have a spinor index under
this $SU(2)$ and another spinor index under the $SU(2)$ R-symmetry
of the 3+1D matrix quiver theory. It should now be the case that there
is a full $SU(4)$ symmetry mixing all four indices.

% --------------------------------------------------------------------- %
% 1+1D SYM
% --------------------------------------------------------------------- %
\subsec{1+1D SYM}

The lowest dimension where a DLCQ description exists is 1+1D.
Compactifying the $U(k)$ 5+1D $(2,0)$ field theory on
$\MT{4}$ gives a 1+1D theory which flows in the IR to
the orbifold $(\MR{5}\times\MT{3})^k/S_k$.  In a limit, we recover
the orbifold $(\MR{8})^k/S_k$ which describes the IR limit of 1+1D SYM.

The spacetime SYM theory has a DLCQ Wilson line as before, which we conjecture
flows
in the IR to a particular $S_k$ element in the longitudinal direction.
To describe the DLCQ sector for the free orbifold $(\MR{8})^k/S_k$
with $\pV = N$,  we need to specify this
$S_k$ element.
This discrete Wilson line is specified by the number of
cycles each of length,
$$
l_1\dots l_k,\qquad \sum j l_j = k.
$$
In the 1+1D IR limit, all the particles are either
left-moving or right-moving at the speed of light.
We will see that the matrix description in terms of the moduli
space ${\cal M}_{k,N}$ of $U(k)$ instantons on $\MT{4}$
can reproduce the multiplicity of states of the, for example, right-moving
modes of the orbifold $(\MR{8})^k/S_k$ with the discrete
$S_k$ Wilson line $\sigma$ given by a single cycle of length $k$.
To see this, we note that since there are no transverse directions
in 1+1D all states in the IR limit have DLCQ energy of precisely
$N$. This means
that to count the number of states in the 1+1D IR limit, we need to
count the number of vacua in the quantum mechanics on ${\cal M}_{k,N}$.
The result of this computation can be predicted by U-duality and there is some
supporting evidence for the prediction \rVafIOD.
We can summarize the result for the cohomology of $\MT{4}$ which contains 8
even-dimensional elements and 8 odd-dimensional elements in the following way:
we
associate a bosonic field $\phi^a_{-n}$  to the
even-dimensional elements ($a=1\dots 8$) and a fermionic field
$\psi^b_{-n}$ to the odd-dimensional elements ($b=1\dots 8$)
where  $n=1,2,\dots$ is the oscillator number.
The cohomology of the resolution of ${\cal M}_{k,N}$
is conjectured to be given by the states of the Fock space generated by
$\phi^a_{-n},\psi^b_{-n}$ at level $N k$.
This agrees with the degeneracy of right-moving states of
$(\MR{8})^k/S_k$ with the special Wilson line $\sigma$ which is
generated by 8 oscillators of fractional $\pV = {j\over k}$
($j=1,2,\dots$) which sum up to $N$.

\newsec{Concluding Comments}

Among the cases we have described is 3+1D SYM compactified on $\MT{2}$.
This is effectively a 1+1D theory and as such has no moduli space. In
the decompactification limit, when $\MT{2} \r \MR{3,1}$,
we also have to specify the point in the moduli space.
There are two qualitative choices: the first leads to a free theory
in the IR, while the second leads to an interacting theory.
In the first case, we can have particle states and so with a matrix
description, we can try to study the scattering of charged particles.
We would hope to obtain a non-perturbative prescription for computing
S-matrix elements involving photons, W-bosons and dyons. However, as
we have seen, it is easiest to find a matrix model
for the conformal theory.

How do we introduce a VEV for the decompactified limit of SYM? In the DLCQ,
the VEV should be an external parameter. We can first consider the
quantum mechanics describing
the $(2,0)$ theory \rquantumfive, which contains a $U(N)$ vector multiplet, an
adjoint hypermultiplet and $k$ fundamental hypermultiplets. Separating the
longitudinal five-branes corresponds to giving bare masses $m_a$,
$a=1,\ldots,k$ to the fundamental hypermultiplets. On taking the quantum
mechanical coupling to infinity, the quartic potential terms still dominate
so we are constrained to the moduli space of instantons, ${\cal M}_{N,k}$.
The mass perturbations then descend to potentials on the moduli space
of instantons. However, for the 3+1D theories the story seems to be
more complicated. From the Coulomb branch perspective, we need
to resolve the $A_{k-1}$ singularity and this leads to problems which
seem related to the issues raised in \rDOS.

We can also try to construct similar models for theories with
less supersymmetry. There are several possible approaches: one way
SYM theories with $\SUSY{2}$ in 3+1D have
been realized is in terms of the low-energy degrees of freedom of
five-branes wrapped on a Riemann surface $\Sigma\subset \MR{3}
\times\MS{1}$ in the limit $\MS{1} \r 0$ \refs{\rmayr, \rWitBR}.
We can ask if there is a matrix model description for this
construction. Let us view this as a limit of a longitudinal
five-brane wrapped on
$\MT{2}\times\Sigma\subset \MT{2}\times\MT{4}$.
This makes the five-brane finite in the transverse directions. If we
 understood more fully the  matrix model for M theory on
$\MT{6}$, we could have
identified this configuration as the sector with the appropriate
quantum number specifying wrapped longitudinal five-branes.
Nevertheless, we can assume that at low-energies, the matrix model
looks like 6+1D SYM. Since we need the limit of $\MT{6}\rightarrow 0$,
the matrix model will be formulated on the dual, large $\MHT{6}$.
The relevant quantum numbers for longitudinal five-branes
are encoded in the Pontryagin class $\Tr F\wdg F$.
Therefore, we are looking for low-energy solutions
of 6+1D SYM with a given Pontryagin class, a point which has also been
discussed in \rGukov.

To get the limit of 3+1D SYM, we have to take
$\MT{6} = \MT{2}\times\MWT{4}$ where $\MT{2}$ is the transverse space
and $\MWT{4}$ is auxiliary and very small.
The matrix model is formulated on the dual $\MHT{2}\times\MHWT{4}$
and as in section (3), we may reduce to the moduli space of holomorphic
maps from $\MT{2}$ to a divisor of $\MHWT{4}$. This means
that for each point $z$ of transverse space $\MT{2}$ there is
Riemann surface
$\Sigma_z\subset\MHWT{4}$ which varies holomorphically with $\MT{2}$.
When the transverse space is decompactified to $\MR{2}$,
the boundary conditions on the Riemann surface are such that
$\Sigma_z$ has to go over to the Seiberg-Witten curve of the theory
as $z\rightarrow \infty$.
On top of that we also need to specify a point on the Jacobian
of each Riemann surface $\Sigma_z$, i.e. $g$ points on $\Sigma_z$
where $g$ is the genus of the surface. Alternatively, $g$ is also
the number of $U(1)$-factors at a generic point in the low-energy theory.
The boundary conditions on these holomorphically varying $g$ points
are such that as $z\rightarrow\infty$, they become electric and
magnetic Wilson lines of the $U(1)^g$ low-energy photons
 along the longitudinal direction. Note that such Wilson lines
are naturally identified with a point on the Jacobian. There are a number
of problems with this possibility so the above comments should be taken as
speculative. In a similar spirit, perhaps matrix models for
$\SUSY{1}$ theories might be constructed starting from
the five-brane constructions of \rneqone.

There is a second approach for obtaining matrix formulations of finite
3+1D $\SUSY{2}$ theories. We can start with the theories described in \rkin\
compactified on $\MT{2}$. Matrix models for these cases can be obtained by
studying a system of 4-branes and 0-branes on $ALE$ singularities. These
models will have $(0,4)$ supersymmetry, and the Coulomb branches of these
models should again describe the conformal points of the SYM theories.

It is interesting to note that we have described SYM theories in terms
of quantum mechanics. While describing the low-energy dynamics of SYM on
a lattice has proven difficult because of chiral fermions, a numerical
analysis of the quantum mechanics should be considerably simpler. After all,
fermions in the quantum mechanics are simply large matrices. Lastly, it
seems possible that some insight into the large $N$ limit for 3+1D SYM
can be obtained by studying how the curves that appeared in that matrix
formulation behave in the large $N$ limit. This might well be the case for
the `t Hooft scaling limit, rather than the large $N$ limit need for
matrix theory.

\bigbreak\bigskip\bigskip\centerline{{\bf Acknowledgements}}\nobreak

It is our pleasure to thank A. Kapustin, S. Ramgoolam,
N. Seiberg and E. Witten
for helpful discussions. The work of O.G. is supported by a Robert H.
Dicke fellowship and by DOE grant DE-FG02-91ER40671; that of S.S. by NSF
grant DMS--9627351.

\listrefs

\end